\documentclass[12pt,english]{article} 
\usepackage{geometry}
\usepackage{graphicx}
\usepackage{color,framed}
\usepackage{tikz}

\usetikzlibrary{calc}

\usepackage{amsmath}
\usepackage{dsfont}
\usepackage{amssymb}
\usepackage{bbm}
\usetikzlibrary{angles, quotes}
\usetikzlibrary{patterns, shapes.geometric, decorations.pathreplacing}
\usepackage{pgfplots}
\usetikzlibrary{arrows.meta}

\usepackage{amsmath}

\newcommand{\lambdabar}{{\mkern0.75mu\mathchar '26\mkern -9.75mu\lambda}}

\begin{document}


 \title {On the reality of quantum states: A pedagogic survey from classical to quantum mechanics} 
 

\author{Moncy V. John \\ School of Pure and Applied Physics, Mahatma Gandhi University, \\ Kottayam, India}

\maketitle


\begin{abstract}
Some recent experiments claim to show that  any model in which a quantum state represents mere information about an underlying physical reality of the system must make predictions which contradict those of quantum theory.  The present work undertakes to investigate the issue of reality, treading a more fundamental route  from the Hamilton-Jacobi equation of  classical mechanics  to the Schrodinger equation of quantum mechanics. Motivation for this is a similar approach from the eikonal equation in geometrical optics to the wave equation in electromagnetic theory.  We rewrite the classical Hamilton-Jacobi equation  as a wave equation and  seek to generalise de Broglie's wave particle duality by demanding that both particle and light waves have the freedom of being described by any  square-integrable function.  This generalisation, which allows superposition  also for matter wave functions, helps us  to  obtain the Schrodinger equation, whose solution  can be seen to be as much objective as the classical mechanics wave function.  Several other equations which one writes in quantum mechanics, including the eigenvalue equations for observables, series expansion of energy states in terms of eigenstates of observables other than energy,  etc.,  can  be written in the classical case too.  Absence of any collapse of the wave function,  entanglement, etc. in the classical realm have their origin in the nonlinearity of the classical wave equation.  These considerations indicate that many of the puzzles in quantum mechanics are present also in classical mechanics in a dormant form, which fact shall help to demystify quantum mechanics to a great extent.

\end{abstract}


\section{Introduction}
Erwin Schrodinger, inspired by Louis de Broglie's principle of wave-particle duality, formulated his famous quantum mechanical wave equation in the year 1926. The  wave function $\Psi$, which is the solution of Schrodinger's wave equation for the system under consideration, was postulated to describe the state of the system at any time. Ever since,  experimental results underscored that $\Psi$ contains all available information regarding the system at a particular time and hence this postulate  is at the heart of quantum theory \cite{cohentann,sakurai}. 

In the early days of quantum mechanics, when one could  do experiments only on systems of large number of particles (an  ensemble), most physicists considered $\Psi$ to be representing our state of knowledge about the system. Thus its use was limited to the computation of expectation values  from repeated measurements by means of the Born probability axiom. Hence $\Psi$ was also called the `probability amplitude'. Even Schrodinger himself was inclined not to accept $\Psi$ as having any reality of higher degree than this. The other major contributors of quantum mechanics, namely, Werner Heisenberg and  Niels Bohr  got around the wave-particle duality of  de Broglie by adopting the position that any classical way of understanding the world would be limited and one can only have knowledge of the outcomes of experiments performed on physical systems.
Heisenberg, in his Copenhagen interpretation of quantum mechanics, argued vehemently against mental pictures of the microworld, such as waves and point particles, and considered the state of the system as an abstract mathematical object (a vector in a linear vector space called Hilbert space). He considered his `state vector' merely as representing the experimenter's knowledge or information about some aspect of reality. Bohr, however, was prepared to concede that an object varied from acting like a particle and like a wave, and also that the wave and particle aspects of the object are complementary and cannot be exhibited at the same time. This proposition later came to be known as Bohr's complementarity principle.   However, it may be noted that for the past one hundred years, the framework of quantum mechanics that every practitioner of physics leans on works exceptionally well in all experimental situations, irrespective of its various interpretations.
 
 With the advancement of modern experimental and observational technology, with which one can now do experiments even on single atoms or photons, the `state of knowledge' view is generally disfavoured and the reality of wave function has gained more acceptance. A recent experimental result \cite{onthereality}  claims to show that any model in which a quantum state represents mere information about an underlying state of the system makes predictions which contradict quantum theory. This has rekindled the age-old debate on the reality of quantum waves
and has attracted lot of interest into  topics that share their borders with philosophy. 

True to the method of science, the best option to look for solutions of such riddles would be to turn to the fundamentals of the theory. 
 In this work, we investigate the issue of  whether the quantum wave function is real (i.e, whether it exists independent of the observer) or whether it represents only the state of knowledge of the observer, by going back to the origin of quantum mechanics.   It may be recalled that until the end of 19th century, classical mechanics \cite{goldstein3} was thought to provide a complete picture, with only a few loose ends to sort out. We now see that fresh outlooks on the reality of quantum wave function are possible by inspecting the direct route from classical mechanics to quantum mechanics. 

In the first part, we take a closer look into the connection between  the eikonal equation, which serves as the basis of geometrical optics,  and the electromagnetic wave equations, which form the basis of wave optics \cite{bornwolf}. The latter equations, due to James Clark Maxwell, were epoch making in the sense that the whole of electric and magnetic phenomena were unified due to it \cite{jackson}. They  provided a general and complete picture from which geometrical optics follows as a limiting case.   A key ingradient in this general formalism is the superposition principle, that is valid for the electromagnetic wave equations but not for the eikonal equations. As a result, an electromagnetic signal can take any square integrable  functional form, while being also the solution of the Maxwell equations. In the second part, we start from the well-known Hamilton-Jacobi (HJ) equation in classical mechanics, which is nonlinear and not obeying  superposition principle. We note that the classical HJ equation can be written in the form of a wave equation (whose solution may be called a classical mechanics wave function) and that this wave equation is also nonlinear, not allowing superposition.  It is now proposed that de Broglie's principle of wave-particle duality  be generalised in such a way that both photon and particle wave functions can be any  function belonging to the set of square integrable functions of position coordinates, which also must  be a solution of their fundamental equation. This  in turn, means that the particle wave equation also must allow the superposition principle.    This helps to modify the HJ equation (or, equivalently, the classical mechanics wave equation) and obtain the required fundamental equation, which we see as the Schrodinger equation in quantum mechanics. In the reverse order, we also note that one can cast the Schrodinger equation in the form of a quantum Hamilton-Jacobi equation, that very much resembles the classical HJ equation \cite{goldstein2}. The former equation can be seen to reduce to the latter one, in the limit when Planck's constant is treated as very small.  

Several other equations, such as the momentum and position eigenvalue equation that one encounters in quantum mechanics, can be written in classical mechanics too. One can assign probabilities for the position of a particle (Born's axiom) and similarly probabilities for each eigenvalue of the observables, just as in quantum mechanics.  The intrinsic spin of fundamental particles are often cited as a property devoid of any classical analogue. But one can see that the  feature that gives rise to intrinsic spin for particles is that their wave functions have more than one component, which is  not altogether strange. All these clearly tell that what  distinguishes quantum mechanics from classical mechanics is  the principle of superposition.  We can see that most puzzles of quantum mechanics are already present   in classical mechanics in a dormant form, which fact demystifies quantum mechanics to a great extent. In particular, we observe that there is no ground for worry on the reality of the wave function.

Most textbooks on quantum mechanics start with the concept of a Hilbert space, where state of the system under consideration is a vector. An advantage of the Hilbert space formalism is that the basic property of quantum systems,  which makes the superposition of their states also  a possible state, is naturally incorporated into the theory. Moreover, the formalism and its compact notations help to track the calculations elegantly and accurately. However,  treating quantum states as abstract vectors in  a linear vector space has the disadvantage that the whole discipline appears esoteric;  as one dealing with mathematical objects with bare connection to reality, except at the final stage of computing experimental outcomes. This disadvantage is further aggravated when quantum mechanics is presented as a disparate discipline, having nominal connection with classical mechanics. 

The present work intends to present quantum mechanics as a natural development of classical mechanics, hoping that this will help students appreciate the discipline with clarity.  This approach alleviates  those several puzzles in quantum mechanics, by showing that the seeds of such puzzles are already there in classical mechanics.

\section{Geometrical optics and wave optics}

\subsection{Eikonal equation in geometrical optics}

Fermat's principle in optics \cite{bornwolf} helps  to identify the path taken by a ray of light  while traveling from one point to another through a medium where the refractive index is $n(x,y,z)$. This principle is stated in terms of  an integral that gives the optical length $\Lambda$ as

\begin{equation}
\Lambda =\int_A^B n(x,y,z)\; ds, \label{eq:opt_length_ch5}
\end{equation}
where $ds$ is the infinitesimal length of a line element  on the path. The actual path taken by the ray of light is that along which this integral is an extremum; i.e., $\delta \Lambda =0$.

William Rowan Hamilton  discovered a partial differential equation whose solution describes the path of a ray of light that passes through an optical system. This equation was later rediscovered by the German mathematician Heinrich Bruns, who named it as  {\sl eikonal equation} (also known as `ray-tracing equation') of  geometrical optics \cite{bornwolf}, and is given by

\begin{equation}
\left(\nabla \Lambda \right)^2=n^2(\bf {r}). \label{eq:eikonal_1}
\end{equation}
 Note that this is a nonlinear differential equation in $\Lambda$ and that it helps to trace all the paths  a ray of light may take as it passes through an inhomogeneous medium. On solving this equation, an indefinite integral $\Lambda$ may be obtained. We  may note that it will appear  similar to the definite integral in equation (\ref{eq:opt_length_ch5}), though not the same as this.

  In the following, we show that the eikonal equation can be obtained from the more general electromagnetic  wave equation. The eikonal can be found to be  a limiting case of the latter, under certain conditions. This implies that geometrical optics follows from wave optics when those conditions are valid. To see this, we start from  Maxwell's  equation written in terms of  the scalar potential. (It can  be derived using the components of its  vector fields as well.) The differential  equation for the scalar field is

\begin{equation}
\nabla ^2\phi =\frac{1}{v^2}\frac{\partial^2\phi}{\partial t^2}=\frac{n^2({\bf r},t)}{c^2}\frac{\partial^2\phi}{\partial t^2},\label{eq:waveqn_3d}
\end{equation}
where $v=v({\bf r},t)$ is the speed of light in the medium and the refractive index $n({\bf r},t)=c/v(\bf {r},t)$ with $c$,  the speed of light in vacuum. Note that this is a linear differential equation in $\phi$.

If the refractive index  is a function of ${\bf r}$ only, i.e., when $n=n({\bf r})$,  one can  separate this equation in terms of its variables ${\bf r}$ and $t$, by assuming

\begin{equation}
\phi({\bf r},t)=u({\bf r})f(t). \label{eq:var_sep}
\end{equation}
The  resulting equations, with the separation constant written as $-1/\lambdabar^2$, are

\begin{equation}
\frac{d^2f}{dt^2}=-\frac{c^2}{\lambdabar^2}f(t), \label{time_part}
\end{equation}
and

\begin{equation}
\nabla^2u({\bf r})=-\frac{n^2({\bf r})}{\lambdabar^2}u({\bf r}). \label{eq:u}
\end{equation}
The first one has the solution 

\begin{equation}
f(t)=\exp(\pm ict/\lambdabar). \label{eq:f}
\end{equation}
In a special case of  refractive index $n=1$ (vacuum), the second equation has the solution 

\begin{equation}
u({\bf r})=e^{ i{\bf k.r}}, \label{eq:wavesoln0}
\end{equation}
where $|{\bf k}|=1/\lambdabar$. If we identify $\lambdabar\equiv \lambda /(2\pi)$, this leads to the usual expression $k=2\pi /\lambda$. For the general case, we shall assume the space part $u({\bf r})$ (of the scalar potential $\phi$) to take the form

\begin{equation}
u({\bf r})=N\exp \left(\frac{i\Lambda({\bf r})}{\lambdabar}\right), \label{eq:wavesoln}
\end{equation}
where  $\Lambda({\bf r})$ and $\lambdabar$ have dimensions of length. Also  $N$ is some appropriate constant of the dimension of $\phi$. In this case, we can express the product function $\phi$  in equation (\ref{eq:var_sep}) as

\begin{equation}
\phi({\bf r},t)=u({\bf r})f(t) =N\exp \left(\frac{i[\Lambda({\bf r})\pm ct]}{\lambdabar}\right). \label{eq:var_sep1}
\end{equation}
We have noted in the above that  $\lambdabar$ has dimensions of length, though its value is not yet specified. In fact, its value can be determined only from the boundary conditions used while solving equation (\ref{eq:u}). It can be seen from this equation that for vacuum  (where $n=1$), 

\begin{equation}
\phi({\bf r},t)=N e^ { i ( {\bf k.r} \pm  \omega t)} \label{eq:wavesoln_1d}
\end{equation}
where  $\omega=2\pi c/\lambdabar$ and $\lambdabar$ is the wavelength of the plane electromagnetic wave in vacuum.

 Using the assumed form in equation (\ref{eq:wavesoln}), the linear differential equation (\ref{eq:u}) for $u$ can be rewritten in terms of $\Lambda$ as 

\begin{equation}
[\nabla \Lambda({\bf r})]^2-n^2({\bf r}) =  i\lambdabar \nabla^2\Lambda({\bf r}). \label{eq:em_eikonal}
\end{equation}
Note that this is the same linear differential equation (\ref{eq:u}) and  there were no approximations till now. But if we restrict ourselves to the case of very small $\lambdabar$ (i.e., $\lambdabar \rightarrow 0$), then the term on the right hand side (second derivative term)  is negligible with respect to the others. Then   the above differential equation (\ref{eq:em_eikonal}) reduces to the  nonlinear eikonal equation (\ref{eq:eikonal_1}). 

What we have proved in this case is that  geometrical optics, characterised by the eikonal equation, is a limiting case (when $\lambdabar \rightarrow 0$) of wave optics (described by the electromagnetic wave equation). One can see that in the case of the plane wave solution obtained for the $n=1$ case mentioned above, the condition that the wavelength $\lambdabar $ is very small is equivalent to the condition that the  frequency of the plane wave is very large.    In the  general case, one can see that the condition to attain geometrical optics limit is that $\lambdabar$ is very small when compared with the scale of variation of $n$.    We may call the full equation (\ref{eq:em_eikonal})  the {\sl electromagnetic eikonal equation} for time-independent refractive index.

As noted above, there is some resemblance between  $\Lambda$ in equation (\ref{eq:opt_length_ch5}), which is the integral involved in the statement  of Fermat's principle, and $\Lambda$ in the solution of equation (\ref{eq:eikonal_1}) or equation (\ref{eq:em_eikonal}). The difference is  that the former is a definite integral, integrated between two given points in space, whereas the latter is just a  function of space coordinates.

\subsection{Eikonal equation for time-dependent refractive index} \label{subsec:eik_time}

Let us now attempt to extend the eikonal equation to  the case of time-dependent refractive index $n(\bf {r},t)$. For this, we have to assume a general form for the solution $\phi({\bf r},t)$,  instead of equation (\ref{eq:var_sep1}), as

\begin{equation}
\phi({\bf r},t)={\cal N} e^{i\xi ({\bf r},t)/\lambdabar}. \label{eq:phixi}
\end{equation}
 Here  $\xi ({\bf r},t)$ is a  new function having dimensions of length and ${\cal N}$ is some appropriate constant having the dimension of $\phi$. As in the previous case, let $\lambdabar$ be a constant  having the dimension of length.  ($\lambdabar$ serves to make the phase dimensionless.)  Using this form for $\phi$,  wave equation (\ref{eq:waveqn_3d}) can be written in terms of $\xi$ as

\begin{equation}
(\nabla \xi )^2-\frac{n^2}{c^2}\left(\frac{\partial \xi}{\partial t}\right)^2 =i\lambdabar \left( \nabla^2 \xi-\frac{n^2}{c^2}\frac{\partial^2 \xi}{\partial t^2}\right)  \label{eq:em_timedep_eikonal}
\end{equation}
This equation is a generalisation of (\ref{eq:em_eikonal}) and   may be  called the {\it  electromagnetic eikonal equation}  for time-dependent refractive index.
In the limit   of $\lambdabar \rightarrow 0$, we may have the right hand side of this equation negligible. In this limiting case,  the  equation becomes

\begin{equation}
(\nabla \xi )^2=\frac{n^2}{c^2}\left(\frac{\partial \xi}{\partial t}\right)^2 \label{eq:timedep_eikonal}
\end{equation} 
This may be called the {\sl time-dependent eikonal equation}, which is the counterpart  of (\ref{eq:eikonal_1})    when the refractive index of the medium is time-dependent. It may be noted that equation  (\ref{eq:em_timedep_eikonal}) is the same electromagnetic wave equation whereas (\ref{eq:timedep_eikonal}) is a nonlinear differential equation.  Here $\Lambda({\bf r})$ is replaced with $\xi({\bf r},t)$.  We note that when $n$ is a function of ${\bf r}$ only, this equation is separable and we may end up with the same eikonal equation (\ref{eq:eikonal_1}).

To rewrite the reduced equation (\ref{eq:timedep_eikonal}) back into the form of an electromagnetic wave equation   in the limit  $\lambdabar \rightarrow 0$ (the geometric optics limit), we may define a new function

\begin{equation}
\Phi({\bf r},t)={\cal N}e^{i\xi({\bf r},t)/\lambdabar} \label{eq:Phi}
\end{equation}
 and use it in  the above equation (\ref{eq:timedep_eikonal}). This results in 

\begin{equation}
(\nabla \Phi )^2=\frac{n^2}{c^2}\left(\frac{\partial \Phi}{\partial t}\right)^2, \label{eq:timedep_eik_phi}
\end{equation} 
having the same form as that in (\ref{eq:timedep_eikonal}), but in terms of $\Phi$. We may consider this as an electromagnetic wave equation in the limit of geometric optics and may call it the {\sl geometrical optics wave equation}. This  is a nonlinear differential equation, unlike the case of the original electromagnetic wave equation (\ref{eq:waveqn_3d}).

\subsubsection{Superposition principle in wave optics}

An important property of the electromagnetic wave equation (\ref{eq:waveqn_3d}) is that it allows superposition of solutions such as that in (\ref{eq:wavesoln_1d}). For example,  consider the one-dimensional case of equation (\ref{eq:waveqn_3d}), where we take $n=1$. The two wave functions $\phi_1=A\exp(k_1x-\omega_1t)$ and $\phi_2=B\exp(k_2x-\omega_2t)$, with $ k_1v=\omega_1$ and $k_2v=\omega_2$, are solutions of this equation.  (These are two solutions that correspond to two different values of the separation constant, namely $\lambdabar_1=1/k_1=\lambda_1/2\pi$ and $\lambdabar_2=1/k_2=\lambda_2/2\pi$.)  Then  the superposition of the two waves, given by $c_1 \phi_1 +c_2 \phi_2$, where $c_1$ and $c_2$ are two constant coefficients, is also a solution to that equation. 

The superposition of solutions is the most important feature that helps in obtaining solutions of  partial differential equations that correspond to particular initial conditions. In the present case, the separation of variables is done to find a general solution to the wave equation of the type (\ref{eq:waveqn_3d}), which can be of the form

\begin{equation}
\phi ({\bf r},t)= \sum_i c_i \exp\left( \frac{i[\Lambda_i (\bf {r})\pm ct]}{\lambdabar_i} \right)
\end{equation}
where $c_i$ can be found from the initial conditions. For vacuum, when the refractive index $n=1$, the above summation series is

\begin{equation}
\phi({\bf r},t) = \sum_i c_i   e^ { i ( {\bf k_i.r} \pm  \omega_i t)} \label{eq:sup_wavesoln_1d}
\end{equation}
This is of the form of a Fourier series. 
The most important feature we note is that superposition principle allows  an electromagnetic signal to take any desired form that obeys the Dirichlet conditions.

 On the other hand,  the eikonal equations (\ref{eq:eikonal_1}) or (\ref{eq:timedep_eik_phi}), obtainable from the original wave equation in the limiting case of $\lambdabar \rightarrow 0$,  are nonlinear differential equations and  do not permit superposition of solutions.

\medskip

{\bf Time-independent refractive index $n({\bf r})$}

\begin{tabular} {|c| c| c|} \hline \label{table:table1}
{\bf Electromagnetic } && {\bf Electromagnetic} \\
{\bf wave equation for} $n=n({\bf r})$  && {\bf eikonal equation} \\
  $\nabla ^2\phi =\frac{n^2({\bf r})}{c^2}\frac{\partial^2\phi}{\partial t^2}$  &$ \Longrightarrow  \phi({\bf r},t)={\cal N} e^{i\Lambda ({\bf r})/\lambdabar}f(t)  \Longrightarrow$ & $[\nabla \Lambda({\bf r})]^2-n^2({\bf r})=i\lambdabar \nabla^2\Lambda({\bf r})$  \\
  && $\frac{d^2f}{dt^2}=-\frac{c^2}{\lambdabar^2}f(t)$ \\ \hline
 && \\
$\Uparrow$ && $\Downarrow \lambdabar \rightarrow 0$ \\
&& \\ \hline  
{\bf Geometrical optics } && {\bf Eikonal equation}  \\
{\bf wave equation} && (Geometrical Optics) \\
$(\nabla \Phi )^2=\frac{n^2}{c^2}\left(\frac{\partial \Phi}{\partial t}\right)^2$ &$ \Longleftarrow  \Phi({\bf r},t)={\cal N} e^{i\Lambda ({\bf r})/\lambdabar}f(t)  \Longleftarrow$ & $[\nabla \Lambda({\bf r})]^2=n^2({\bf r})$ \\ 
&&  $\frac{d^2f}{dt^2}=-\frac{c^2}{\lambdabar^2}f(t)$ \\ \hline
\end{tabular}

Table  1.  Top row corresponds to the full electromagnetic equations and bottom row to equations in the geometrical optics limit. Left column corresponds to wave equations and the right column to eikonal equations.

\medskip

{\bf Time-dependent refractive index $n({\bf r},t)$}

\begin{tabular} {|c| c| c|} \hline \label{table:table2}
{\bf Electromagnetic } & & {\bf Electromagnetic} \\
{\bf wave equation for}  $n=n({\bf r},t)$  && {\bf eikonal equation} \\
 $\nabla ^2\phi =\frac{n^2({\bf r},t)}{c^2}\frac{\partial^2\phi}{\partial t^2}$  & $ \Longrightarrow  \phi({\bf r},t)={\cal N} e^{i\xi ({\bf r},t)/\lambdabar} \Longrightarrow$ &  \\
 && $(\nabla \xi )^2-\frac{n^2}{c^2}\left(\frac{\partial \xi}{\partial t}\right)^2=i\lambdabar \left( \nabla^2 \xi-\frac{n^2}{c^2}\frac{\partial^2 \xi}{\partial t^2}\right)$ \\ \hline
 && \\
$\Uparrow$ && $\Downarrow \lambdabar \rightarrow 0$ \\
&& \\ \hline
{\bf Geometrical optics}  && {\bf Time-dependent} \\
 {\bf wave equation} && {\bf eikonal equation} \\
    & $ \Longleftarrow  \Phi({\bf r},t)={\cal N} e^{i\xi ({\bf r},t)/\lambdabar} \Longleftarrow$ & (Geometrical Optics) \\
   $(\nabla \Phi )^2=\frac{n^2}{c^2}\left(\frac{\partial \Phi}{\partial t}\right)^2$ &&  $(\nabla \xi )^2=\frac{n^2}{c^2}\left(\frac{\partial \xi}{\partial t}\right)^2$ \\ \hline
  \end{tabular}
  
Table 2. Top row corresponds to the full electromagnetic equations and bottom row to equations in the geometrical optics limit. Left column corresponds to wave equations and the right column to eikonal equations.

\section{Comparison with mechanics}

\subsection{Hamilton-Jacobi theory}

In the  general case  of a time-dependent Hamiltonian in classical mechanics, one writes  the HJ equation \cite{goldstein3} as

\begin{equation}
H\left( q,\frac{\partial S}{\partial q},t\right)+\frac{\partial S}{\partial t}=0.
\end{equation}
where $S$ is the Hamilton's principal function. This  is in general an arbitrary function of $q^i$ and $t$ with constant parameters  $\alpha^i$.  In cases where  the  Hamiltonian $H$ does not depend explicitly on time,  the HJ equation can be written as

\begin{equation}
\frac{\partial S}{\partial t}=-H\left(q,\frac{\partial S}{\partial q}\right) =-a. \label{eq:partialS}
\end{equation}
where $a$ is some constant. The last step follows from the fact that  the Hamiltonian in this case is a constant of motion. Integrating this equation gives 

\begin{equation}
S(q,\alpha,t)= -a t+W(q,\alpha),\label{eq:S}
\end{equation} 
where the time-independent $W(q,\alpha)$ is called the Hamilton's characteristic function. This relation connects Hamilton's principal and characteristic functions.

The restricted (time-independent) HJ equation for Hamilton's characteristic function $W$, which  is a function of $n$ generalised coordinates $q^i$ and the $n$ constants of integration $\alpha^i$; i.e., $W=W(q^1,q^2,...q^n,\alpha^1, \alpha^2, ..\alpha^n)\equiv W(q,\alpha)$ is of the form

\begin{equation}
H\left( q, \frac{\partial W}{\partial q}\right) =a ,
\end{equation} 
where $a$ is one of the integration constants $\alpha^i$ or some combination of them. This equation is valid when the Hamiltonian is independent of time $t$ and is a constant of motion.
 For cases in which  the kinetic energy T is purely quadratic in the velocity components and the potential $V$ is independent of velocity, we can identify $a=E$, the total energy of the system.  For such case  of a single particle in 3-dimensional space, one can write the above equation as

\begin{equation}
\frac{1}{2m}\left(\nabla W\right)^2+V({\bf r})=E. \label{eq:HJ_char_eqn2}
\end{equation}
or 

\begin{equation}
\left(\nabla W\right)^2 -{2m[E-V({\bf r})]}=0. \label{eq:HJ_char_eqn3}
\end{equation}
We observe that this equation is quite similar to the eikonal equation in  (\ref{eq:eikonal_1}), with $W$ playing the role of the optical distance $\Lambda$ and $\sqrt{2m[E-V({\bf r})]}$ that of the refractive index $n$. In fact, this similarity is very deep-rooted and is the key to the connection between quantum and classical laws of mechanics.

 In the following, let us consider  the special form of the HJ equation, which is valid in cases where the kinetic energy contains only terms quadratic in generalised velocities and  the potential $V$ is  time-dependent. In this case, we have

\begin{equation}
\frac{1}{2m}(\nabla S)^2+V({\bf r},t)+\frac{\partial S}{\partial t}=0. \label{eq:HJ_eqn}
\end{equation} 
 This HJ equation in classical mechanics can be considered as the counterpart of the time-dependent eikonal equation   for geometrical optics, as given in (\ref{eq:timedep_eikonal}). The above cases of restricted HJ equation (\ref{eq:HJ_char_eqn3}) and the HJ equation (\ref{eq:HJ_eqn}) are clearly nonlinear, due to the presence of the terms with $(\nabla W)^2$ and $(\nabla S)^2$, respectively. 
 
\subsection{Classical mechanics wave equation} \label{subsec:HJ_waveqn}
 
 Having observed the  resemblance of  geometrical optics with the mechanics of point particles, we proceed further to explore this similarity to its full extent.  For this, we first rewrite the above HJ equation (\ref{eq:HJ_eqn}) in terms of a new function defined by

\begin{equation}
X ({\bf r},t)\equiv e^{iS({\bf r},t)/\hbar}, \label{eq:psi_S}
\end{equation}
where $\hbar$ is a constant having the dimensions of action $S$. The resulting equation is of the form

\begin{equation}
-\frac{\hbar^2}{2m}\frac{1}{X}(\nabla X)^2 +V({\bf r},t)X =i\hbar \frac{\partial X}{\partial t} \label{eq:class_waveqn}
\end{equation}
This  may be called a {\sl classical mechanics wave equation}. As stated above, the optical counterpart of this  is the geometrical optics wave equation (\ref{eq:timedep_eik_phi}). We shall refer $X({\bf r},t)$ a {\sl classical mechanics wave function}.

We must note particularly that the constant $\hbar$  in equation (\ref{eq:psi_S}) is introduced just as a parameter to make the phase of $X$ dimensionless. When the form (\ref{eq:psi_S}) is used in the above equation (\ref{eq:class_waveqn}), we get back the HJ equation (\ref{eq:HJ_eqn}) and the parameter $\hbar$ disappears. Consequently, also the equation of motion that connects the theory to physically measured quantities is independent of the constant $\hbar$ and hence in the present case, it can be any  arbitrary constant. 

In classical mechanics,  Hamilton's principal function $S({\bf r},t)$, defined in the configuration space with specified values of integration constants, gives the dynamics of the classical system, with the help of the HJ equation. In a similar way, one can consider also the above classical mechanics wave function $X({\bf r},t)$ in equation (\ref{eq:psi_S})  as capable of providing  the trajectories of the system.

 As an example of the classical mechanics wave function, let us consider the harmonic oscillator of mass $m$ and angular frequency $\omega$. Hamilton's principal  function in this case is given by equation \cite{goldstein3}
 
 \begin{equation}
S=-E t+\sqrt{2mE}\int dq \sqrt{1-\frac{m\omega^2q^2}{2E}} \label{eq:S_int_ho}
\end{equation}
which can be integrated to get explicitly

\begin{equation}
S=-E t+\frac{E}{\omega} \arcsin\left[ \sqrt{\frac{m\omega^2}{2E}} q\right] +\frac{q}{2}\sqrt{2mE -m^2\omega^2q^2} +C  \label{eq:S_ho}
\end{equation}
 The corresponding classical mechanics wave function is 
 
\begin{equation}
X({q},t)=e^{iS({q},t)/\hbar}=A e^{-iE t/\hbar}e^{\frac{i}{\hbar}\left[ \frac{E}{\omega} \arcsin\left( \sqrt{\frac{m\omega^2}{2E}} q\right) +\frac{q}{2}\sqrt{2mE -m^2\omega^2q^2} \right]}  \label{eq:X_ho}
\end{equation} 
 where $A=e^{iC/\hbar}$.
 
 Though  the already known expression for $S$ is  used in   the above equation to obtain $X({q},t)$, one can directly start with the equation (\ref{eq:class_waveqn}) for the solution of mechanical problems. In this case, one attempts to solve  (\ref{eq:class_waveqn}) to get $X({\bf r},t)$ with appropriate boundary conditions, and then identify the Hamilton principal  function $S$ from it. The usual methods of obtaining the trajectories in the HJ formalism can then  be employed without  modifications.

 \section{Superposition in mechanics} \label{sec:sup_pos}

Let us now look closely into the issue of superposition in classical HJ theory. As we have seen earlier, when the kinetic energy $T$ is purely quadratic in velocity components and the potential is independent of velocity, the HJ equation can be written in the form of equation (\ref{eq:class_waveqn}), which we have referred to as the classical mechanics wave equation. When $V=V({\bf r})$, the Hamiltonian is a constant of motion and we can separate this equation into space and time parts by assuming a solution of the form 
\begin{equation}
X ({\bf r},t)=\chi({\bf r}) f(t) .
\end{equation}
In this case, let us denote the separation constant as $E$. It is easy to see that the space part  of the classical mechanics wave equation is

\begin{equation}
-\frac{\hbar^2}{2m}\frac{1}{\chi_i({\bf r})}[\nabla \chi_i({\bf r})]^2+V({\bf r})\chi_i({\bf r})=E_i\chi_i({\bf r}). \label{eq:HJchar_waveqn}
\end{equation}
This is not a linear differential equation, since the first term contains the square of the derivative  of $\chi$, divided by $\chi$. Using $\chi_i({\bf r})=\exp[iW_{E_i}({\bf r})/\hbar]$  into this equation,  one finds that it is equivalent to the restricted (time-independent)  HJ equation (\ref{eq:HJ_char_eqn2}). 
It can be easily seen that the time-dependent part of the classical mechanics wave equation is similar to equation (\ref{time_part}) and has the solution

\begin{equation}
f_i(t)=e^{-iE_it/\hbar}. \label{eq:time_part}
\end{equation}
Together with this, the classical mechanics wave function can be written as

\begin{equation}
X_i ({\bf r},t)=\chi_i({\bf r}) f_i(t) =e^{i[W_{E_i}({\bf r})-E_it]/\hbar }. \label{eq:Psi_E}
\end{equation}

Using this, one can see that  the  differential equations (\ref{eq:class_waveqn}) and (\ref{eq:HJchar_waveqn}) are equivalent to the HJ equation and the restricted HJ
equation, respectively. The solutions of the latter equations  shall be independent of $\hbar$ and hence in this discussion, $\hbar$ can take any arbitrary value.

A  comparison of the HJ theory with the case of optics can now be made.
We note that both the geometrical optics wave equation  (\ref{eq:timedep_eik_phi}) and  the  classical mechanics  wave equation (\ref{eq:class_waveqn})   are nonlinear differential wave equations, which do not allow the superposition principle. For instance,  even when $X_1$ and $X_2$ are two solutions of the  classical mechanics wave functions  that correspond to energies $E_1$ and $E_2$, a linear combination $c_1 X_1 +c_2 X_2$ will not be a solution to it. In general, a  linear combination of such products in the form

\begin{equation}
\sum_{i} c_i X_{i}({\bf r},t), \label{eq:Xi_superposition}
\end{equation}
  cannot be a general solution to  (\ref{eq:class_waveqn}).
This case is similar to the geometrical optics wave equation (\ref{eq:timedep_eik_phi}). In fact, this mathematical result  is related to an important physical property of the classical systems that they can be  described only by  certain definite functions obeying appropriate boundary conditions. An arbitrary function cannot describe  the state of the system, as discussed in the following subsection.

\subsection{Physical meaning of the absence of superposition}

The HJ  function $S(q,\alpha,t)$, which is a solution of the HJ  equation obtained for the set of integration constants $\{ \alpha^i \}$, can  be said to  represent a state of the system. Equivalently we can say that the classical mechanics wave function $X(\bf {r},t)$, which is the solution of the corresponding wave equation (\ref{eq:class_waveqn}),  represent the same state of the system.  Let us call such states as the  {\sl energy state} of the system. We use the term `energy state'  to distinguish it from the conventional `state' of the system represented by a point in phase space. In Newtonian  mechanics, specifying the location of a point in phase space requires information regarding the values of all generalised coordinates and generalised momenta of the system at a given time. On the other hand,   with the help of the energy state, one can draw only a  set of trajectories $q^i(\alpha , \beta , t)$ in the position (configuration) space for the system.  Earlier it was  pointed out that one of the integration constants $\alpha^i$ or a particular combination of them can be identified as the energy $E$ of the system. Hence we shall  refer the state corresponding to the wave function $X$, obtained as a solution to the equation (\ref{eq:class_waveqn}),  as  representing an {\sl energy state} of the system.

The nonvalidity of superposition principle  for the classical mechanics wave equation indicates that in classical mechanics,  a physical system  can remain only  in any one energy state at a time. When a system is  described by $X({\bf r})$,  it has a well-defined energy, which is also  a constant of motion. Suppose that $X_1({\bf r})$ and $X_2({\bf r})$ are two such solutions of the equation that correspond to two different  energies $E_1$ and $E_2$. If the linear combination $c_1 X_1 +c_2 X_2$ can also be a solution, that would mean that the system  corresponds to two different energies simultaneously. The absence of superposition in classical HJ theory can be understood as  indicating that  such a situation is forbidden.

Equally important is the fact that in the classical case, the state of a system cannot be described by an arbitrary (wave) function. This is in sharp contrast with the case of wave optics, where superposition of waves  allow an electromagnetic signal to take  arbitrary functional forms.

\section{Solving the mechanical problem in HJ theory} \label{subsec:soln_mech_HJ_alt}

We now  discuss an approach  used to solve the mechanical problem in the HJ theory. The attempt is to show that the equation

\begin{equation}
p^i=\frac{\partial S(q,\alpha,t)}{\partial q^i}, \label{eq:canon2_1_2}
\end{equation} 
 can be used to solve for the trajectories, by direct integration. To use this as an equation of motion, one must have the canonical momentum $p^i$  expressed in terms of $q^i$, $\dot{q}^i$ and $t$. This can be done using
 
\begin{equation}
p^i=\frac{\partial L}{\partial \dot{q}^i}, \label{eq:canonp}
\end{equation} 
 where $L$ is the Lagrangian of the problem.   (It must be kept in mind that the variable $p^i$ in this equation is not always the mechanical momentum. However, in such cases, one can find relations connecting these two.)
  
As an example for this method, let us solve the harmonic oscillator problem by directly integrating equation (\ref{eq:canon2_1_2}), when the complete solution $S$ of the HJ equation for this case is available. Thus in this approach, we first solve the HJ equation to obtain $S$, as in equation (\ref{eq:S_int_ho}) or (\ref{eq:S_ho}). The equation of motion (\ref{eq:canon2_1_2}) for the harmonic oscillator can be written by combining

\begin{equation}
p=\sqrt{2mE-m^2\omega^2q^2}.
\end{equation}
and the equation (\ref{eq:canonp}), which gives the canonical momentum as 

$$
p=\frac{\partial L}{\partial \dot{q}} =m\dot{q}.
$$
We thus get 

$$
m\dot{q}=\sqrt{2mE-m^2\omega^2q^2}
$$
or

\begin{equation}
\dot{q}^2+\omega^2q^2=\frac{2E}{m}.
\end{equation}

Integrating this with respect to time gives the solution as
\begin{equation}
q=A \sin(\omega t+\epsilon)
\end{equation}
where $A$, the amplitude of oscillation is given by $\sqrt{2E/m\omega^2}$. 

  \subsection{Energy state and trajectories}
   The energy state of the system can be represented using trajectories, once we obtain $X({\bf r})$. Such solutions correspond to the set of values  $\alpha^i$.  Then one  obtains the Hamilton principal function $S$ using the form (\ref{eq:psi_S})  as

\begin{equation}
S=\frac{\hbar}{i}\log X.
\end{equation}
The  trajectories, which represent the energy state, can now be found by following the HJ method. Various trajectories $q^i(\alpha^i,\beta^i,t)$ can be drawn starting from an ensemble of initial points in this space.

 \subsection{Equation of motion and the momentum eigenfunction} \label{subsec:eqn_mtn}

Once  the classical mechanics wave function in  the form (\ref{eq:Psi_E}) is available with us, the classical trajectories can be obtained by using the equations of motion, as described above. In three-dimensions, this latter approach relies on integrating the equation of motion (\ref{eq:canon2_1_2}) as

\begin{equation}
{\bf p}=\nabla S = \nabla W = \frac{\hbar}{i}\frac{1}{X}\nabla X =\frac{\hbar}{i}\frac{1}{\chi({\bf r})}\nabla \chi({\bf r}). \label{eq:eqn_of_motion}
\end{equation}
Here $\bf {p}$ is the canonical momentum of a single particle in 3-dimensional Cartesian coordinates. This equation of motion can now be rewritten  in the form of a first order differential equation

\begin{equation}
-i\hbar \nabla \chi ({\bf r})={\bf p}({\bf r}) \chi({\bf r}). \label{eq:eqn_of_motn}
\end{equation}
In the special case of a free particle, where ${\bf p}$ is a constant vector, we  write this  equation by replacing $\chi(\bf{r})$ with $u(\bf {r})$ as

\begin{equation}
-i\hbar \nabla u ({\bf r})={\bf p} \; u({\bf r}). \label{eq:mom_eqn}
\end{equation}
The solution of  this equation has the form

\begin{equation}
u({\bf r})= {\cal N} e^{i{{\bf p}}.{\bf r}/\hbar} \label{eq:mom_state}
\end{equation}
where  ${\cal N}$ is some constant.  Equation (\ref{eq:mom_eqn}) has the form of an eigenvalue equation, corresponding to a differential operator $-i\hbar \nabla$. We may now call  $u({\bf r})$, which is the eigenfunction of $-i\hbar \nabla$,  a {\sl momentum  eigenfunction} corresponding to the eigenvalue ${\bf p}$.  A general solution of the classical mechanics wave equation (\ref{eq:HJchar_waveqn})  for systems under any non-zero potentials  can be written  as linear combinations of these momentum eigenfunctions $u({\bf r})$.  In cases where the system is bounded in a finite volume or has periodic boundary conditions, the momentum eigenvalues will be discrete, which may be denoted as ${\bf p}_i$. We can then expand $\chi({\bf r})$ as a summation series, which in fact is  the Fourier series expansion for $\chi({\bf r})$ :

\begin{equation}
\chi({\bf r})=\sum_{i} c_{i} \; u_{i}   =\sum_{i} c  _{i} \;  e^{i{{\bf p}_i}.{\bf r}/\hbar}. \label{eq:expansion_chi}
\end{equation}
Here, $c_{i}$ are appropriate coefficients in the expansion. Summation is indicated here, assuming that ${{\bf p}_i}$ takes discrete values. Here we are expanding $\chi({\bf r})$  in terms of the basis functions $u_{i}({\bf r})$, which  represent  various momentum states of the system. 

In the general case of an unbound one-particle system, in which ${\bf p}$ is a continuous variable,  the summation in the above expression must be replaced with an integration, so that one can write

\begin{equation}
\chi({\bf r})=\int c({\bf p})e^{i\frac{{\bf p}.{\bf r}}{\hbar}}d^3{\bf p}.
\end{equation}
This is the familiar Fourier (inverse) transform of a function $c({\bf p})$ to obtain $\chi({\bf r})$. We shall discuss these relations further in the following sections.


\section{Quantum Mechanics - Schrodinger wave equation} \label{sec:SE}
Where did we get that (equation) from? Nowhere. It is not possible to derive it from anything you know. It came out of the mind of Schrödinger.

\noindent Richard Feynman

Isaac Newton's corpuscular theory of light explained several phenomena in optics, such as reflection, refraction, dispersion, etc. (This topic is is now referred to as geometrical optics.) But for the explanation of other optical phenomena such as interference, diffraction, etc., Christiaan Huygens'  wave theory of light was needed.   In 1905,  Albert Einstein used Planck's notion of `quantum of radiation energy' to explain photoelectric effect and this led to the concept of what is now called photons. Einstein conceived the quantum idea as providing  particle nature for radiation. Apparently, he  hoped to revive Newton's corpuscular picture for light, in some new form. In 1909, he even attempted to derive a law of motion for the quantum of energy of electromagnetic radiation, but did not develop the theory further. In 1923, Luis de Broglie  postulated that just like photons,  matter particles  may also possess dual nature and hence may have  wave properties as well. His hypothesis was that nature loves symmetry and hence  wave-particle duality must be common to both photons and electrons. In his doctoral thesis in 1924, he put forward the idea of phase waves, which really acted as the much awaited trigger for what  later was termed  the  `quantum revolution'. 

Though the subject of quantum mechanics stemmed from  de Broglie's hypothesis of `wave-particle duality', its initial development  was  along seemingly different directions. These ramifications were a result of intensely polarised debates in the 1920's and the new quantum mechanics had to face  many hurdles before getting accepted as  a proper scientific theory. Perhaps this was the most fierce battle in the history of human knowledge, that shook not only the conceptual foundations of mechanics, but the very fundamental world view of mankind. The aftershocks of quantum theory have not yet subsided.

 A couple of years after  de Broglie's path-breaking idea, a formulation of   quantum mechanics named  {\sl matrix mechanics} was proposed by Werner Heisenberg, Max Born  and Pascual Jordan.  Almost around the same time, Erwin Schrodinger discovered his {wave  equation}.  Inspired by de Broglie's hypothesis of phase waves,  Schrodinger   tried to find a wave equation for an electron in the hydrogen atom. Denoting  de Broglie's phase wave as $\Psi =\Psi({\bf r},t)$,  Schrodinger at first attempted to use the standard wave equation   

\begin{equation}
\nabla ^2\Psi =\frac{1}{v_p^2}\frac{\partial^2\Psi}{\partial t^2},\label{eq:waveqn_SE}
\end{equation}
for a one-particle system, with $v_p$ as the phase velocity of the particle.  However, when applied to the hydrogen atom  to predict its energy levels, this equation disagreed with experiments and he abandoned it.(This equation is now known as Schrodinger's original relativistic equation.) It was only later that he attempted a nonrelativistic version of it, which is the now famous {\sl Schrodinger equation}

\begin{equation}
-\frac{\hbar^2}{2m}\nabla^2 \Psi({\bf r},t) +V({\bf r},t)\Psi({\bf r},t)=i\hbar \frac{\partial \Psi({\bf r},t)}{\partial t}. \label{eq:SE}
\end{equation}
Here $V({\bf r},t)$ is the potential in which the particle moves. Schrodinger found that the new mechanics based on this equation agrees exceptionally well with the observed features of the hydrogen spectrum. 

  According to standard quantum theory, the solution  $\Psi$ to the Schrodinger equation, which is called the quantum wave function,  represents the `state' of the system. It is usually stressed that $\Psi$ contains all the available information regarding the system.

  \subsection{Quantum Hamilton-Jacobi equation}
  
We have mentioned above that the Schrodinger equation assumes a pivotal role in the accurate description of the microworld.  Let us now move one step backward and write an equation equivalent to the Schrodinger equation, by using $\Psi=\exp(i{S}/\hbar)$ in equation (\ref{eq:SE}). The result is

\begin{equation}
\frac{1}{2m}(\nabla {S})^2+V({\bf r},t)+\frac{\partial {S}}{\partial t}=\frac{i\hbar}{2m}\nabla^2 {S}. \label{eq:mod_HJ_eqn}
\end{equation} 
This equation appears similar to the HJ equation  (\ref{eq:HJ_eqn}), except for its nonzero right hand side. This is often referred to as the {\sl quantum HJ equation}. We shall see later that contrary to the HJ  equation, the above quantum HJ equation and also its solutions always involve the constant $\hbar$.   One can  find the value of this constant only from experiments. In cases where we have the term with $\hbar$ quite negligible when compared to other terms in the equation, the classical HJ equation can be regained. This is  the classical limit of quantum mechanics.

\section{Alternative approach to obtain the Schrodinger equation}
It was mentioned above, quoting Feynman,  that the Schrodinger equation cannot be derived from anything; it was the genius of Schrodinger which helped to arrive at the equation. However, some heuristic arguments are often put forward in support of this equation. Here we present an alternative argument based on the superposition principle, that helps to deduce it.

We have seen that Maxwell's electromagnetic wave equation  (\ref{eq:waveqn_3d}) is linear and  obeys the superposition principle, whereas  the eikonal equation of geometrical optics is nonlinear, and hence does not obey it. It was also noted that  when we take the limit $\lambdabar \rightarrow 0$  in the  electromagnetic eikonal equations (\ref{eq:em_eikonal}) or (\ref{eq:em_timedep_eikonal}), they tend to the eikonal equation in geometrical optics, making  the superposition principle  invalid. Along with this, it was   seen that in classical mechanics, the HJ equation  and the equivalent classical mechanics wave equation (\ref{eq:class_waveqn}) are nonlinear and hence do not have the advantage of   superposition of energy states as their solutions. The import of this is that there is a discrepancy between photon wave functions and particle wave functions; the former can take any square integrable functional form, which is also a solution of Maxwell equations, whereas the latter cannot have such a freedom in the classical HJ formalism. 
 It can now be argued that the de Broglie's wave particle duality  between photons and particles may be extended to a deeper level: Since photons can be described by a  wave function which can be the superposition of solutions of Maxwell's wave equation, matter particles  must also be described  by a wave function, which can be the solution of a linear  wave equation allowing superposition of energy states.  Below we show that this symmetry consideration helps in arriving at the Schrodinger equation.

 For this, we must seek whether   a generalisation, such as that from the nonlinear eikonal equation in geometrical optics to the more general linear electromagnetic wave equation, is possible  for classical HJ theory. In other words, we ask whether it is possible to show that the classical mechanics wave equation (\ref{eq:class_waveqn}) arises as a limiting case of a (quantum) wave equation that obeys the superposition principle.     We shall see later that the reason behind those several puzzles in quantum mechanics is the introduction of this superposition principle.

The first step  towards this is the  identification that the kinetic energy term in the classical mechanics wave equation (\ref{eq:class_waveqn})   is
$-({\hbar^2}/{2m}) ({1}/{X^2})(\nabla X)^2$.
In fact, this is  the same term $(\nabla S)^2/2m$ in the HJ equation, which causes the nonlinearity in the  classical mechanics wave equation. We note that the equation (\ref{eq:class_waveqn}) can be made linear by modifying the kinetic energy term by adding a term $\frac{\hbar^2}{2m}\nabla \left(\frac{1}{X}\nabla X\right)$ to it; i.e.,

\begin{equation}
\hbox{Classical} \; K.E. = -\frac{\hbar^2}{2m} \left[ \frac{1}{X}(\nabla X)\right]^2 \rightarrow  -\frac{\hbar^2}{2m}  \left[ \frac{1}{X}(\nabla X)\right]^2  + \frac{\hbar^2}{2m} \nabla \left( \frac{1}{X} \nabla X \right) .  \label{eq:alt_derivation_SE}
\end{equation}
Addition of this term effectively leads to a zero-point energy in quantum mechanics. 
Making the corresponding change in equation (\ref{eq:class_waveqn}) and relabeling $X$ as $\Psi$, we obtain the second order linear Schrodinger differential equation (\ref{eq:SE}).

 Schrodinger's quantm mechanical wave equation and the equivalent quantum HJ equation are shown in the first row of the following table. In the limit of $\hbar \rightarrow 0$, the latter goes over to the classical HJ equation. This, in turn, is equivalent to the classical mechanics wave equation, both of which are shown in the second row. By making the latter equation linear by the addition of a term as in equation (\ref{eq:alt_derivation_SE}), we regain  Schrodinger's quantum wave equation.
 
{\bf Time-dependent potential $V=V({\bf r},t)$}

\begin{tabular} {|c| c| c|} \hline \label{table:table3}
&& \\
 {\bf Quantum mechanics} & &  {\bf Quantum} \\
{\bf wave equation for}  &&  {\bf Hamilton-Jacobi equation} \\
 $V=V({\bf r},t)$  & $ \Longrightarrow  \Psi({\bf r},t)={\cal N} e^{i\hat{S} ({\bf r},t)/\hbar} \Longrightarrow$ &  \\
$ -\frac{\hbar^2}{2m}\nabla^2 \Psi +V({\bf r},t)\Psi=i\hbar \frac{\partial \Psi}{\partial t}$   && $\frac{1}{2m}(\nabla \hat{S})^2+V({\bf r},t)+\frac{\partial \hat{S}}{\partial t}=\frac{i\hbar}{2m}\nabla^2 \hat{S}      $ \\ 
&& \\ \hline
&& \\
$\Uparrow$ && $\Downarrow \hbar \rightarrow 0$ \\
&& \\ \hline
&& \\
 {\bf Classical mechanics} && {\bf Hamilton-Jacobi equation} \\
 {\bf wave equation}    & $ \Longleftarrow  X({\bf r},t)={\cal N} e^{iS ({\bf r},t)/\hbar} \Longleftarrow$ & (Classical Mechanics) \\
$-\frac{\hbar^2}{2m}\frac{1}{X}(\nabla X)^2 +V({\bf r},t)X =i\hbar \frac{\partial X}{\partial t}$             &&   $\frac{1}{2m}(\nabla S)^2+V({\bf r},t)+\frac{\partial S}{\partial t}=0$    \\
  && \\ \hline
  \end{tabular}
  
Table 3.  Top row corresponds to quantum mechanical equations and bottom row to classical mechanical equations. Left column corresponds to wave equations and the right column to HJ-type trajectory equations.

\section{Time-independent potentials} \label{sec:timeind_pot}

In  cases where the potential $V$ is independent of $t$, one can expect to separate the Schrodinger  equation into its time and space parts. As in the earlier case of the electromagnetic wave equation (\ref{eq:time_part}), we can separate the Schrodinger equation into its space and time parts by assuming $\Psi({\bf r},t)=\psi({\bf r})f(t)$. The time-part of this equation is

$$
i\hbar \frac{df}{dt}=Ef(t),
$$
which has the solution

$$
f(t)\propto e^{-iEt/\hbar},
$$
with the separation constant as $E$.  The above product wave function can then be written in the form

\begin{equation}
\Psi_E({\bf r},t)= \psi({\bf r},E)e^{-iEt/\hbar}, \label{eq:Psi_psi}
\end{equation}
 Here $\psi({\bf r},E)$ is the solution to the space part of the equation, given by

\begin{equation}
-\frac{\hbar^2}{2m}\nabla^2 \psi({\bf r},E) +V({\bf r})\psi({\bf r},E)= E  \psi({\bf r},E). \label{eq:timeind_SE}
\end{equation}

 This is referred to as the {\it time-independent Schrodinger equation}, which has the form of  an eigenvalue equation. 
 
 Equations corresponding to those in the above table, written for the case of time-independent potentials $V({\bf r})$, is given in the following.

{\bf Time-independent potential $V=V({\bf r})$}

\begin{tabular} {|c| c| c|} \hline \label{table:table4}
&& \\
 {\bf Quantum mechanics}  &&  {\bf Time-independent quantum} \\
 {\bf wave equation for} && {\bf Hamilton-Jacobi equation} \\
  $V=V({\bf r})$ &$ \Longrightarrow  \Psi({\bf r},t)={\cal N} e^{i\hat{W} ({\bf r})/\hbar}f(t)  \Longrightarrow$ & $-i\hbar \nabla^2\hat{W}({\bf r}) +[\nabla \hat{W}({\bf r})]^2=2m(E-V)$ \\
 $ -\frac{\hbar^2}{2m}\nabla^2 \Psi +V({\bf r})\Psi=i\hbar \frac{\partial \Psi}{\partial t}$   && $i\hbar \frac{df}{dt}=Ef(t)$ \\ 
 && \\ \hline
 && \\
$\Uparrow$ && $\Downarrow \hbar \rightarrow 0$ \\
&& \\ \hline  
&& \\
{\bf Classical mechanics } &&  {\bf Time-independent (restricted)} \\
{\bf wave equation}&& {\bf Hamilton-Jacobi equation} \\
&& (Classical Mechanics) \\
  $-\frac{\hbar^2}{2m}\frac{1}{X}(\nabla X)^2 +V({\bf r})X =i\hbar \frac{\partial X}{\partial t}$                        &$ \Longleftarrow  X({\bf r},t)={\cal N} e^{iW ({\bf r})/\hbar}f(t)  \Longleftarrow$ & $(\nabla W)^2=2m(E-V)$ \\ 
&&  $    i\hbar \frac{df}{dt}=Ef(t)         $ \\ 
&& \\ \hline
\end{tabular}

Table 4. Top row corresponds to the quantum mechanical equations and bottom row to classical mechanical equations. Left column corresponds to wave equations and the right column to HJ-type trajectory equations.

\section{Momentum and  Hamiltonian  operators} \label{sec:mom_Hamil_op}
Let us denote a differential operator, such as  $\frac{d}{dx}$, $\frac{d^2}{dx^2}$, etc., or some combinations of them, by the symbol ${\cal L}_{op}$. An eigenvalue equation for this operator can be written in the general form  

\begin{equation}
{\cal L}_{op} y_i({\bf r}) = \kappa_i \; y_i({\bf r}), \label{eq:eigen}
\end{equation}
where $\kappa_i$ is the eigenvalue and $y_i$, the corresponding eigenfunction. In this section,  we  assume that the eigenvalue $\kappa_i$ has discrete values and $i$ is the integer that labels the eigenvalue. The case with continuous eigenvalues shall be discussed in the following sections.

\subsection{Momentum operator} \label{subsec:mom_oper}

 Recall that we obtained equation  (\ref{eq:mom_eqn}) as a special case of   the classical equation of motion (\ref{eq:eqn_of_motion})  and wrote it in the form of an  eigenvalue equation

\begin{equation}
-i\hbar \nabla u ({\bf r},{\bf p})={\bf p} \; u({\bf r},{\bf p}). \label{eq:mom_eqn1}
\end{equation}
Here ${\bf p}$ is some constant vector having dimensions of momentum and is the eigenvalue.  In this equation, $-i\hbar \nabla$ is a differential operator 

\begin{equation}
{\bf p}_{op} \equiv  -i \hbar \nabla , \label{eq:mom_op}
\end{equation}
which is called the momentum operator and $u({\bf r},{\bf p})$ are its eigen functions. We  now write the eigenvalue equation as 

$${\bf p}_{op}u ({\bf r},{\bf p})={\bf p} \; u({\bf r},{\bf p}).$$ 
As seen from equation (\ref{eq:mom_state}), the eigenstate of momentum operator that corresponds to the eigenvalue ${\bf p}$ is

\begin{equation}
u({\bf r},{\bf p})= {\cal N} e^{i{{\bf p}}.{\bf r}/\hbar}. \label{eq:mom_state3}
\end{equation}
Here ${\cal N}$ is a constant. As an extension of the discussion in Subsec. \ref{subsec:eqn_mtn}, one notes that any quantum mechanical wave function can be expanded into a Fourier series in terms of the eigenfunctions of the momentum operators.

\subsection{Hamiltonian operator} \label{subsec:energy_oper}

The time-independent Schrodinger equation (\ref{eq:timeind_SE}) can also be considered  as an eigenvalue equation. In this case,   the differential operator 

\begin{equation}
H_{op}\equiv  -\frac{\hbar^2}{2m}\nabla^2  +V({\bf r})  \label{eq:H_op}
\end{equation}
may be used to write  equation (\ref{eq:timeind_SE}) as

\begin{equation}
H_{op}\psi_E({\bf r},E) =E\psi_E({\bf r},E). \label{eq:timeind_SE1}
\end{equation}
Here the energy $E$ is the eigenvalue and $\psi_E({\bf r},E)$ is the corresponding eigenfunction. Since $E$ can be identified as the energy of the system, $H_{op}$ is called the {\it Hamiltonian operator}. (It may be noted that at present, there is no other reason behind this nomenclature. We shall see later that $H_{op}$ is a `Hermitian operator' and hence will have real  eigenvalues  $E$.)  With this form of $H_{op}$, the Schrodinger equation (\ref{eq:SE}) can be written as

\begin{equation}
H_{op}\Psi({\bf r},t)=i\hbar \frac{\partial \Psi({\bf r},t)}{\partial t}. \label{eq:SE1}
\end{equation}

While solving equation (\ref{eq:timeind_SE1}), we  see that  the energy eigenvalues may turn out to be discrete, depending on the  boundary conditions. Denoting these discrete values as $E_i$,  the eigenvalue equation (\ref{eq:H_op}) is 

\begin{equation}
H_{op}\psi_i({\bf r}) =E_i\psi_i({\bf r}) \label{eq:timeind_SE2}
\end{equation}
Due to the linearity of this equation, any linear combination of eigenfunctions of the form $\psi_i({\bf r})$; i.e.,

\begin{equation}
\psi({\bf r})=\sum_i c_i \; \psi_i({\bf r}), \label{eq:Psi_expansion}
\end{equation}
 will also be its solution. In other words, any solution $\psi({\bf r})$  of the time-independent Schrodinger equation can be expanded into a series of the above form. Again, due to the linearity of equation (\ref{eq:SE}), any general state $\Psi({\bf r},t)$ of the one-particle system  can be expressed as a superposition of energy eigenfunctions

   \begin{equation}
   \Psi({\bf r},t)=\sum_i c_i\; \psi_i({\bf r}) e^{-iE_i t/\hbar}. \label{eq:unit_evol}
   \end{equation}
    It may  be noted that we have used 
    
    \begin{equation}
    \Psi({\bf r},0)\equiv  \psi({\bf r}),
    \end{equation}
as in equation (\ref{eq:Psi_expansion}).     Once the coefficients $c_i$ are found, the future time-evolution of the wave function state is obtained by this expression. The coefficients $c_i$  in equations such as (\ref{eq:Psi_expansion}) can be found by making use of the property called Hermiticity of the Hamiltonian operator. 
    
It must be noted that this is for the case in which the potential is $V({\bf r})$, a function of ${\bf r}$ only.  In this case, since $c_i$ are constants, the above function will continue to be a solution of the Schrodinger equation for all future times, due to its property of linearity.

 For a free particle, $V({\bf r})=0$. In this case, the time-independent Schrodinger equation  has the solution
 
 \begin{equation}
 \psi ({\bf r}) ={\cal N} e^{i{\bf k} .{\bf r}}, \label{eq:freep_wfn}
 \end{equation}
where $|{\bf k}|=\sqrt{2mE}/\hbar$. Here the energy eigenvalue $E$ is a continuous variable. Since this function satisfies the eigenvalue equation 

$${\bf p}_{op} \psi({\bf r})=\hbar {\bf k}\psi({\bf r}),$$
 we note that $\psi({\bf r})$, as in equation (\ref{eq:freep_wfn}), is  the same function  as the momentum eigenfunction $u({\bf r})$ in equation (\ref{eq:mom_state}) with ${\bf p}=\hbar {\bf k}$. Thus in this special case of a free particle,  eigenfunctions of the Hamilton operator are the same as the eigenfunctions of the momentum operator.

 \section{The postulate of probability in quantum mechanics} \label{sec:prob}

\subsection{Born's probability axiom} \label{subsec:born}

The statistical interpretation of the wave function in quantum mechanics was made  by Max Born. We have already mentioned  that  Born was a major contributor to the first formulation of quantum mechanics named `matrix mechanics.  He proposed this,  together with Werner Heisenberg and Pascual Jordan. However, Max Born is primarily recognised for  the interpretation of $|\Psi|^2$ as the probability density, something that he worked on alone and published in 1926. For this work, he was awarded Nobel prize in 1954.  His axiom  closely resembles the one in electromagnetic theory that the intensity of  light wave at a point is proportional to the square of the amplitude of the electromagnetic wave at that point. 

The Born axiom  states that when  a particle is described by a wave function $\Psi({\bf r},t)$ and a measurement of its position  is made, the probability to find the particle in a volume element $dV$ around the point ${\bf r}$ is  $\Psi ^{\star}({\bf r},t)\Psi({\bf r},t)dV=|\Psi({\bf r},t)|^2dV$. Let us denote $|\Psi({\bf r},t)|^2\equiv P({\bf r},t)$, which may be termed the probability density. (In view of this, it is appropriate to call $\Psi({\bf r},t)$ as the probability amplitude.) If a particle is stable and does not disappear in any other way, one can be sure to find it somewhere in space. Thus Born's axiom naturally leads to the conclusion that the  probability to find the particle `anywhere' in  space is  equal to unity,  The condition can now  be written as

\begin{equation}
\int \Psi^{\star}({\bf r},t)\Psi({\bf r},t)\;dV=\int |\Psi({\bf r},t)|^2\;dV =1.\label{eq:normal1}
\end{equation}
When this  is satisfied, $\Psi({\bf r},t)$ is said to be normalised to unity.
One can show, with the help of the Schrodinger equation (\ref{eq:SE1}), that if a wave function is normalised at any given instant $t$, it will stay normalised for all times. For this, we also need the complex conjugate of the Schrodinger equation

\begin{equation}
\left(H_{op}\Psi({\bf r},t)\right)^{\star}=-i\hbar \frac{\partial \Psi^{\star}({\bf r},t)}{\partial t}. \label{eq:SE1_cc}
\end{equation}

The above statement can be proved by showing that the time derivative of $\int \Psi^{\star}({\bf r},t)\Psi({\bf r},t)\;dV$ vanishes; i.e., by showing 

\begin{equation}
\frac{d}{dt} \int \Psi^{\star}({\bf r},t)\Psi({\bf r},t)\;dV =0. \label{eq:cons_totprob}
\end{equation}
Using the Schrodinger equation and its complex conjugate, one can write 

\begin{eqnarray}
\frac{d}{dt} \int \Psi^{\star}({\bf r},t)\Psi({\bf r},t)\;dV &=&\int\left(\frac{\partial\Psi^{\star}}{\partial t}\Psi+\Psi^{\star}\frac{\partial \Psi}{\partial t}\right)dV \\
&=& \frac{1}{i\hbar} \left[\int -(H_{op}\Psi)^{\star}\Psi dV+ \int \Psi^{\star}H_{op}\Psi dV\right] \\
&=&0.
\end{eqnarray}
The last step follows from the property of Hermiticity  of the Hamilton operator. This proves our assertion.

\subsubsection{Expectation (mean) values of position}
The above probability axiom is  directly useful in evaluating the expectation value of position of a particle, during position measurements. Let a large number of measurements be made on an ensemble of identically prepared systems, all in the same state $\Psi({\bf r},t)$. Based on  the above postulate of  probability density, one can  evaluate the expectation or mean value of position, denoted by $\langle {\bf r} \rangle$, as

\begin{equation}
\langle {\bf r} \rangle = \int_V {\bf r}|\Psi({\bf r},t)|^2 dV \label{eq:expctn_position1}
\end{equation}
We shall see later in Section \ref{sec:born_any} that the probability axiom can be extended to other observable physical quantities as well.

     \section{Hermitian operators} 

Linear operators such as the Hamiltonian operator in equation (\ref{eq:H_op}) or the momentum operator in equation (\ref{eq:mom_op})   have the important property called {\sl Hermiticity}. To see what this property is, let us first define the adjoint operator corresponding to  any operator ${\cal L}_{op}$. The adjoint  ${\cal L}_{op}^{\dagger}$ of an operator ${\cal L}_{op}$ is the one which satisfies the condition 

\begin{equation}
\int_{x_1}^{x_2} f^{\star}(x)[{\cal L}_{op}g(x)]dx = \int_{x_1}^{x_2} [{\cal L}^{\dagger}_{op}f(x)]^{\star} g(x)dx.
\end{equation}
Note that the integral is taken over the interval $x_1\leq x\leq x_2$, where the scalar product is defined.
A linear operator ${\cal L}_{op}$ is said to be {\sl Hermitian} (or {\sl self-adjoint}) over this interval, if it is its own adjoint; i.e., if it has the property

\begin{equation}
\int_{x_1}^{x_2} f^{\star}(x)[{\cal L}_{op}g(x)]dx = \int_{x_1}^{x_2} [{\cal L}_{op}f(x)]^{\star} g(x)dx, \label{eq:hermiticity}
\end{equation}
where $f(x)$ and $g(x)$ are any two functions.

\subsection{Reality of eigenvalues and orthogonality of eigenfunctions} \label{subsec:reality_orthogonality} 
 
Now, consider an eigenvalue equation of the form (\ref{eq:eigen}) in one dimension

\begin{equation}
{\cal L}_{op} y_i({ x}) = \kappa_i y_i({x}), \label{eq:eigen1}
\end{equation}
and assume ${\cal L}_{op}$ is Hermitian. 

\noindent {\bf Theorem:} Hermitian operators have the special property that  their eigenvalues are always  real and two eigenfunctions corresponding to distinct eigenvalues of them are orthogonal to each other.

 If any one of the eigenvalues is    {\it g-fold degenerate} (meaning there are $g$ eigenfunctions corresponding to the same eigenvalue), one can always  construct $g$ orthogonal eigenfunctions corresponding to this eigenvalue.  (The procedure used for this is called {\sl Gram-Schmidt orthogonalisation}.)  The orthogonality relation between normalised eigenfunctions $y_i(x)$ and $y_j(x)$  can be written conveniently by the equation

\begin{equation}
\int_{x_1}^{x_2} y_i^{\star}(x)y_j(x) dx=\delta_{ij}, \label{eq:orthogonality}
\end{equation}
 where $\delta_{ij}$ is the Kronecker $\delta$-symbol, defined by

\begin{eqnarray}
\delta_{ij}& = & 1, \qquad \hbox {when}\; i=j \\
            &=& 0, \qquad \hbox {when}\; i\neq j .\\  
\end{eqnarray}
If all the eigenfunctions are  normalised and are orthogonal to each other, we may refer to such a set of eigenfunctions  as an {\sl orthonormal} set.

\subsection{Completeness of eigenfunctions and the expansion postulate} \label{subsec:complete}

   The set of eigenfunctions of a Hermitian operator may be termed a {\it complete  set}, if any square integrable function $\phi(x)$ defined in the given domain of values of $x$ can be expanded as a series in terms of this set of eigenfunctions. This shall be of the form

   \begin{equation}
\phi(x)=\sum_i c_i \; y_i (x). \label{eq:y_expansion}
\end{equation}
It can  now be explained   how one can  find the coefficients $c_i$ in the expansion (\ref{eq:y_expansion}). Multiplying both sides of this equation with $y_j^{\star}(x)$ and integrating over the above interval, we get
   
   \begin{equation}
   c_j=\int_{x_1}^{x_2} y_j^{\star}(x)\phi(x)dx, \label{eq:c_n}
\end{equation}   
making use of    the orthogonality property of $y_i(x)$.

This result helps us to make a formal statement of the {\sl completeness property}  envisaged in equation (\ref{eq:y_expansion}). We rewrite this equation as

\begin{equation}
\phi(x)=\sum_i y_i(x)\int_{x_1}^{x_2} y_i^{\star}(x^{\prime})\phi(x^{\prime})dx^{\prime}=\int_{x_1}^{x_2}\left[\sum_i y_i(x) y_i^{\star}(x^{\prime})\right] \phi(x^{\prime})dx^{\prime}.
\end{equation}
  The expression in square brackets can be identified with the {\it Dirac $\delta$-function}; i.e.,
   
   \begin{equation}
   \sum_i y_i(x)y_i^{\star}(x^{\prime}) =\delta(x-x^{\prime}). \label{eq:complete}
   \end{equation}
    Here the Dirac $\delta$-symbol, which is denoted as  $\delta(x-x^{\prime})$, has the defining  property  
   
\begin{equation}
\int_{x_1}^{x_2} f(x)\delta (x-x^{\prime}) dx = f(x^{\prime}), \label{eq:deltafn}
\end{equation}
 for all functions $f(x)$, provided the limits of integration $[x_1,x_2]$ includes the point $x=x^{\prime}$. Equation (\ref{eq:complete}) is called the completeness property or {\sl closure property} of the eigenfunctions. 
 Only a complete set of basis functions can satisfy the completeness relation.

 
\section{Observables} \label{sec:hermiticity}

 In quantum mechanics, each physical quantity has a corresponding operator. It was stated above that Hermitian operators have real eigenvalues and that its eigenfunctions corresponding to different eigenvalues are orthogonal to each other. A Hermitian operator can be said to represent an {\sl observable} if its orthonormal eigenfunctions form a complete basis to expand a wave function state of the physical system.

  First, let us check whether the   momentum operator ${\bf p}_{op}$ defined by equation (\ref{eq:mom_op}) and  the Hamiltonian operator $H_{op}$ defined  by equation (\ref{eq:H_op})   are  Hermitian and  represent observables.
 
 \subsection{Momentum} \label{subsec:momentum_observable}

 It is easy to see that the momentum operator  ${\bf p}_{op}\equiv -i\hbar \nabla$  satisfies  equation (\ref{eq:hermiticity}) and is hence a Hermitian operator. To show  this, let us consider the simple case where the system is a single particle confined to  the interval $[x_1,x_2]$ in a one-dimensional space. Here its wave functions  vanish for both $x\leq x_1$ and $x\geq x_2$. Let $f(x)$ and $g(x)$ be any two such functions.  Integrating by parts,  the left hand side of equation (\ref{eq:hermiticity}) (with ${\cal L}_{op} ={ p}_{op}=-i\hbar \frac{d}{d x}$) can be found to be 
 
 \begin{eqnarray}
 \int_{x_1}^{x_2} f^{\star}(x)[{ p}_{op}g(x)]dx &=&-i\hbar\int          _{x_1}^{x_2} f^{\star}(x)\left[\frac{d}{d x} g(x)\right]dx \\
 &=& -i\hbar\left\lbrace \left[f^{\star}(x)g(x)\right]_{x_1}^{x_2}-\int \left[\frac{d}{d x} f^{\star}(x)\right] g(x)dx \right\rbrace \\
 &=&\int_{x_2}^{x_2} [{  p}_{op}f(x)]^{\star} g(x)dx, \label{eq:hermiticity_p}
 \end{eqnarray}
which is equal to  its right hand side, showing that $p_{op}$ is Hermitian. 
Here, we made use of the fact that $f(x)$ and $g(x)$  vanish  at $x_1$ and $x_2$. The same result can be obtained in the general case of many particle systems in three dimensions too. Thus ${\bf p}_{op}$ is Hermitian.

As seen in Sec. \ref{sec:mom_Hamil_op},  the eigenstates of ${\bf p}_{op}$ for the discrete case are 

\begin{equation}
u_i({\bf r}) = {\cal N} e^{i{\bf p}_i.{\bf r}/\hbar} . \label{eq:3d_mom_eigfn}
\end{equation}
 Here, ${\cal N}$ is a normalisation constant. The discrete case arises when the particle is confined in a box. Let us consider a one-dimensional box of length $L$, with boundaries at $x=-L/2$ and $x=+L/2$. The momentum eigenfunction, which vanishes at the boundaries, are to be of the form

\begin{equation}
u_{xi}(x)=\frac{1}{\sqrt{L}}e^{ip_{xi} x/\hbar}, \qquad p_{xi}=\frac{2\pi\hbar n_{xi}}{L}, \qquad n_{xi}=0,\pm 1,\pm 2, ... \label{eq:1d_mom_eigfn}
\end{equation}
The orthogonality relation in this case is seen as

\begin{equation}
\int u_{xi}^{\star}(x)u_{xj}(x)dx = \delta_{ij}.
\end{equation}
In equation (\ref{eq:1d_mom_eigfn}), the normalisation factor is chosen as ${\cal N}=\frac{1}{\sqrt{L}}$ to agree with this equation. In the three-dimensional case, we can write the orthogonality relation as

 \begin{equation}
 \int_V u_{i}^{\star}({\bf r})u_{j}({\bf r}) dV=\delta_{ij}.
 \end{equation}
Here one must choose the normalisation factor in equation (\ref{eq:3d_mom_eigfn}) as ${\cal N}=\frac{1}{L^{3/2}}$. 
 
 In the case of a particle occupying the whole of  space, rather than  a box, the momentum eigenvalue ${\bf p}$ is a continuous variable and the  momentum eigenfunction is $u({\bf r},{\bf p})={\cal N} e^{i{\bf p}.{\bf r}/\hbar}$. In this case, the normalisation constant must be chosen as 
 
\begin{equation}
{\cal N}=\frac{1}{(2\pi \hbar)^{3/2}}, \label{eq:norm_N}
\end{equation}
to obtain the orthogonality relation

 \begin{equation}
 \int_V u^{\star}({\bf r},{\bf p})u({\bf r},{ {\bf p}^{\prime}}) dV=\delta ({\bf p}-{\bf  p}^{\prime}).\label{eq:ortho_p}
 \end{equation}
 The functions $u_i({\bf r}) = {\cal N}e^{i{\bf p}_i.{\bf r}/\hbar}$ or $u({\bf r},{\bf p}^{\prime})={\cal N}e^{i{\bf p}^{\prime}.{\bf r}/\hbar}$, with ${\bf p}/\hbar={\bf k}$,  are the base functions used in the Fourier series or Fourier transform, respectively, and hence they are known to form a complete set. This establishes the fact that momentum is an observable, as per the above definition.
 
 We shall now explicitly write the expansion of any square-integrable wave function $\psi({\bf r})$ in the whole of space, in terms of the momentum eigenfunctions, as
 
 \begin{equation}
 \psi({\bf r}) =\int c({\bf p}) u({\bf r, p}) d^3p = \frac{1}{(2\pi\hbar)^{3/2}}\int c({\bf p})e^{i{\bf p.r}/\hbar}d^3p,
 \end{equation}
where the integral is taken over the whole of momentum space.
The coefficients $c({\bf p})$ are obtained as in equation (\ref{eq:c_n}):

\begin{equation}
c({\bf p})= \int u^{\star}({\bf r,p}) \psi({\bf r}) dV =\frac{1}{(2\pi\hbar)^{3/2}}\int \psi({\bf r})e^{-i{\bf p.r}/\hbar}dV,
\end{equation}
where the integral is taken over the whole of configuration space.
In terms of the propagation constant ${\bf k} = {\bf p}/\hbar$, the above equation for $\psi({\bf r})$ is  the  Fourier transform of $c({\bf p})$, which is a well-known result. Hence one can reasonably deduce that the momentum eigenfunctions form a complete set and that momentum is an observable.

 \subsection{Energy} \label{subsec:hamiltonian_observable}

 In a similar way, we can see that the energy of a system,  represented by the Hamiltonian operator,  is an observable. The Hamiltonian operator for a single particle moving in a potential $V({\bf r})$ was identified in equation (\ref{eq:H_op}) as
 
 \begin{equation}
H_{op}\equiv  -\frac{\hbar^2}{2m}\nabla^2  +V({\bf r}).  \label{eq:H_op2}
\end{equation}
To show that this is an observable,  we need  to show that $H_{op}$ is a Hermitian operator and that its eigenfunctions form a complete set. Let us first consider for simplicity a single free particle in one-dimension, which has  $V(x)=0$.   In this case, one can  write equation (\ref{eq:hermiticity}) as 

\begin{eqnarray}
\int_{x_1}^{x_2} f^{\star}(x)H_{op}g(x)dx&=& -\frac{\hbar^2}{2m} \int_{x_1}^{x_2} f^{\star}(x)\frac{d^2}{d x^2} g(x)dx \\
&=& -\frac{\hbar^2}{2m} \int_{x_1}^{x_2} f^{\star}(x)d\left(\frac{d}{d x} g(x)\right) \\
&=&  -\frac{\hbar^2}{2m} \left\lbrace \left[f^{\star}(x)\frac{d}{d x} g(x)\right]_{x_1}^{x_2}-\int_{x_1}^{x_2} \left[ \frac{d}{d x} f^{\star}(x)\right] \left[\frac{d}{d x} g(x)\right] dx \right\rbrace  \\
&=&  \frac{\hbar^2}{2m} \left\lbrace \left[\frac{d f^{\star}(x)}{dx} g(x)\right]_{x_1}^{x_2}-\int_{x_1}^{x_2} \left[\frac{d^2}{d x^2}  f^{\star}(x)\right] g(x)dx \right\rbrace  \\
&=&\int_{x_1}^{x_2}\left[ -\frac{\hbar^2}{2m} \frac{d^2}{d x^2}  f(x)\right]^{\star} g(x)dx .\\
\end{eqnarray}
Here  $f(x)$ and $g(x)$ are any two functions which vanish, both at $x_1$ and $x_2$. This result shows that $H_{op}$ is Hermitian when there is no potential. It is easy to see that $V(x)$, which is  a function of the real variable $x$, is already in a Hermitian form. Hence in the general case also, $H_{op}$ is a Hermitian operator.

The energy eigenvalue equation for the discrete case is  given in equation (\ref{eq:timeind_SE2}). We recall that the index $i$  labels the energy eigenstates with  eigenvalues $E_i$. Here a summation over $i$ implies a summation over the discrete energy levels. Instead, while treating energy as a continuous variable $E$,  we  write  the eigenvalue equation as

\begin{equation}
H_{op}\psi_E({\bf r},E)=E\psi_E({\bf r},E),
\end{equation}
where the parameter $E$ is written  inside the parenthesis. 

 The orthogonality of energy eigenfunctions for  the discrete and continuous cases are written as
 
 \begin{equation}
 \int_V \psi_i^{\star}({\bf r})\psi_j({\bf r})dV=\delta_{ij},
 \end{equation}
and

\begin{equation}
\int_V \psi_E^{\star}({\bf r},E)\psi_E({\bf r},E^{\prime})dV = \delta(E-E^{\prime}), \label{eq:ortho_E}
\end{equation}
respectively.  When the energy eigenstates form a complete set,  the completeness relation (\ref{eq:complete})  can be written for the discrete and continuous cases of energy eigenstates, as

   \begin{equation}
   \sum_i \psi_i({\bf r})\psi_i^{\star}({\bf r}^{\prime}) =\delta({\bf r}-{\bf r}^{\prime}), \label{eq:complete_E_dis}
   \end{equation}
and

\begin{equation}
\int \psi_E({\bf r},E)\psi_E^{\star}({\bf r}^{\prime},E)dE = \delta({\bf r}-{\bf r}^{\prime}), \label{eq:complete_E_cont}
\end{equation}
respectively. When these results hold, we may consider that $H_{op}$ represents an observable, which is the energy of the system. Note that here the completeness of eigenfunctions of $H_{op}$ is only an assumption.

The linear expansion  (\ref{eq:Psi_expansion})  for some wave function $\psi({\bf r})$ may now be written in terms of  eigenfunctions belonging to the continuous energy eigenvalues $E$  as

   \begin{equation}
   \psi({\bf r})=\int c(E) \psi_E({\bf r},E)  dE. \label{eq:expansion_cont}
   \end{equation}
 The coefficients $c(E)$ can be found as discussed above in equation (\ref{eq:c_n}) as
 
 \begin{equation}
 c(E)=\int \psi_E^{\star}({\bf r},E) \psi({\bf r}) dV.
\end{equation}

\subsection{Position} \label{subsec:pos_observable}

In the Hamiltonian formalism of classical mechanics, momentum and position are treated almost equivalently. In the above sections, we have written down the eigenvalue equations for momentum and energy. A similar eigenvalue equation for position may be postulated as 

\begin{equation}
{\bf r}_{op}w ({\bf r},{\bf r}^{\prime})={\bf r}^{\prime} \; w({\bf r},{\bf r}^{\prime}).
\end{equation} 
For the eigenvalues of position to be real, we must assume   ${\bf r}_{op}$ to be  a Hermitian operator. We expect that the position vector ${\bf r}$ is   a continuous variable. In the above equation, the operator ${\bf r}_{op}$ has eigenvalues ${\bf r}^{\prime}$ and eigenfunctions $w({\bf r},{\bf r}^{\prime})$.  Assuming a Hermitian operator, the operator ${\bf r}_{op}$ can  have orthogonal   eigenfunctions, obeying the relations
   
   \begin{equation}
   \int_V w^{\star}({\bf r},{\bf r}^{\prime}) w({\bf r},{\bf r}^{\prime \prime}) dV=\delta({\bf r}^{\prime}-{\bf r}^{\prime\prime}), \label{eq:ortho_x}
   \end{equation}
similar to equations (\ref{eq:ortho_p}) and (\ref{eq:ortho_E}).    
   
   The completeness relation for the position eigenfunctions, in a manner similar to that in equation (\ref{eq:complete_E_cont}), must be of the form
   
   \begin{equation}
\int_V w({\bf r},{\bf r}^{\prime})w^{\star}(\tilde{{\bf r}},{\bf r}^{\prime})dV^{\prime} = \delta({\bf r}-\tilde{{\bf r}}). \label{eq:complete_x}
\end{equation}

When the completeness relation is valid, it should be possible to expand an arbitrary wave function in terms of $w({\bf r},{\bf r}^{\prime})$ as

\begin{equation}
\psi({\bf r})=\int_V c({\bf r^{\prime}}) w({\bf r},{\bf r}^{\prime}) dV^{\prime}.
\end{equation}

 We shall see later that the eigenfunction $w({\bf r},{\bf r}^{\prime})$ of the operator ${\bf r}_{op}$ is the same as  $\delta({\bf r}-{\bf r}^{\prime})$,  the Dirac $\delta$-function. In that case, it is easy to see that the above equations (\ref{eq:ortho_x}) and (\ref{eq:complete_x}) are satisfied, so that $w({\bf r},{\bf r}^{\prime})$ form a complete orthonormal set. Hence position is an observable, according to the above definition.

\section{Operators corresponding to other physical quantities}

We have  seen that to every physical system, there corresponds a wave function  describing the energy state of the system and  this wave function shall be a solution of the Schrodinger equation (\ref{eq:SE}). 
 In the above section, it was shown that for the physical system under consideration, position, momentum and energy have   Hermitian operators ${\bf r}_{op}$, ${\bf p}_{op}$ and $H_{op}$, respectively. The eigenstates of these operators may form a complete basis and in that case, they obey the completeness relation. Hence they can be considered as  `observables' of the physical system. 
 
   Now the question arises whether the eigenstates of  any other operator can also be used as base functions to expand $\Psi({ {\bf r}^{\prime}},0)$. The answer is in the affirmative,  provided its operators are expressed in terms of ${\bf r}_{op}$ and ${\bf p}_{op}$,  they are Hermitian operators and their eigenfunctions form a complete set. It was seen that the Hamiltonian operator $H_{op}$, in the particular form in equation (\ref{eq:H_op}), is Hermitian and a wave function of the form $\Psi({\bf r},t)$ can be written as linear combination of energy eigenstates. A Hermitian operator that corresponds to a physical quantity $A=A({\bf r,p})$ can be constructed using ${\bf r}_{op}$ and ${\bf p}_{op}$ as $A_{op}=A({\bf r}_{op},{\bf p}_{op})$. Further, if they are `observables', meaning if their eigenfunctions form a complete set,  any admissible  quantum wave function $\Psi({\bf r},t)$ of the system can be expanded in terms of such eigenfunctions.

As mentioned above,  one can find operators corresponding to such dynamical variables of the system by replacing the position and momentum variables in it with their respective quantum mechanical operators. This procedure may sometimes involve some operator ordering ambiguities.  When the eigenfunctions of such operators form a complete set, they too are considered as observables.

\section{An example: angular momentum}
As an example of obtaining the operator corresponding to a physical quantity, we shall consider the case of angular momentum. Classically, the angular momentum of a single particle is defined as

\begin{equation}
{\bf L}={\bf r}\times {\bf p}
\end{equation}
where ${\bf r}$ is the coordinate of the particle and ${\bf p}$ is the momentum. The operator corresponding to ${\bf L}$ is chosen as

\begin{equation}
{\bf L}_{op}={\bf r}_{op}\times {\bf p}_{op}=-i\hbar {\bf r}\times \nabla.
\end{equation}
which is a Hermitian operator.
One can construct an operator, which is the square of this, denoted as $L^2_{op}$. Its eigenfunctions are the famous functions called `spherical harmonics'. We encounter these operators in several three-dimensional problems in quantum mechanics.

\subsection{Spin angular momentum and motion in an electromagnetic field}

Consider a particle of mass $m$ and charge $e$ in an electromagnetic field , where the vector potential is represented by ${\bf A}$ and the scalar potential by $\phi$. The classical Hamiltonian  in this  case is 

\begin{equation}
H=\frac{1}{2m} \left( {\bf p} - \frac{e}{c}{\bf A}({\bf r},t)\right)^2+e\phi({\bf r},t)
\end{equation} 
The Schrodinger equation for this problem is written as

\begin{equation}
i\hbar \frac{\partial }{\partial t} \Psi = \left[ \frac{1}{2m}\left(-i\hbar \nabla -\frac{e}{c}{\bf A}\right)^2+e\phi \right]\Psi
\end{equation}

Now consider the case of an electron. It is known from  experiments that the electron  possesses an (internal) angular momentum (spin), whose components in an arbitrarily chosen direction are only the values $+\hbar/2$ and $-\hbar/2$. Many other elementary particles do also have non-zero spin. Those with above value as half-integral multiples of $\hbar$ are called fermions and those with integral multiples of $\hbar$ are called bosons.

It was Wolfgang Pauli who found that if the wave function of  a Schrodinger equation has two components, written in a column matrix form, 

\begin{equation}
\Psi=\left(\begin{array}{c}
  \Psi_{+} \\
  \Psi_{-} \end{array}\right),
\end{equation} 
the effect of a magnetic moment 

\begin{equation}
\mu_{s} = g \frac{e}{2mc}{\bf S}
\end{equation}
corresponding to a spin angular momentum ${\bf S}$ can be accounted for. Here $g$ is called Lande $g$-factor. His version of the Schrodinger equation, called the Pauli equation, can be written as

\begin{equation}
i\hbar \frac{\partial}{\partial t} \left(\begin{array}{c}
  \Psi_{+} \\
  \Psi_{-} \end{array}\right)=\left[ \left(\frac{1}{2m}\left(-i\hbar \nabla - \frac{e}{c}{\bf A}({\bf r},t)\right)^2 +e\phi ({\bf r},t)\right) {\bf 1} +\mu_s\; {\bf \sigma }{\bf . B}\right]  \left(\begin{array}{c}
  \Psi_{+} \\
  \Psi_{-} \end{array}\right) \label{eq:pauli_eqn}
\end{equation}
where ${\bf \sigma}$ corresponds to the three Pauli matrices $ {\bf \sigma}=\sigma_1,\; \sigma_2\; \sigma_3$, given as

\begin{equation}
 \sigma_1=
  \left( \begin{array}{cc}
  0 & 1  \\
  1 & 0   
   \end{array}  \right) \\[11pt] \qquad 
 \sigma_2=
  \left( \begin{array}{cc}
  0 & -i  \\
  i & 0   
   \end{array}  \right) \hbox{and} \qquad 
   \sigma_3=
  \left( \begin{array}{cc}
  1 & 0  \\
  0 & -1   
   \end{array}  \right)  
    ,\\[11pt]
\end{equation} 
and ${\bf 1}$ refers to a $2\times 2$  identity matrix. Equation (\ref{eq:pauli_eqn}), in fact, refers to the two equations
\begin{equation}
i\hbar \frac{\partial \Psi_{+}}{\partial t}= \left[ \left(\frac{1}{2m}\left(-i\hbar \nabla - \frac{e}{c}{\bf A}({\bf r},t)\right)^2 +e\phi ({\bf r},t)\right)\right] \Psi_{+} +\mu_s[B_z \Psi_{+}+(B_x-iB_y)\Psi_{-}] \label{eq:pauli1}
\end{equation}

and
\begin{equation}
i\hbar \frac{\partial \Psi_{-}}{\partial t}= \left[ \left(\frac{1}{2m}\left(-i\hbar \nabla - \frac{e}{c}{\bf A}({\bf r},t)\right)^2 +e\phi ({\bf r},t)\right)\right] \Psi_{-} +\mu_s[(B_x + iB_y)\Psi_{+}-B_z\Psi_{-}] \label{eq:pauli2}
\end{equation}

The above Pauli formalism of extending  the wave function state of the electron to one with  two components, which feature effectively appears as the intrinsic spin of the electron  interacting with the magnetic field, is said to have no classical analogue. Absence of classical analogue in this case must be understood as the absence of more-than-one component wave functions in the classical cases.  Even in the quantum realm, particles which do not have any intrinsic spin will have a wave function with only one component, as can be seen directly from equation (\ref{eq:pauli_eqn}). In this case where $\mu_s=0$, the two equations (\ref{eq:pauli1}) and (\ref{eq:pauli2}) reduce to a single equation, and there is only  one component   for the wave function. It is be noted that since $\mu_s $ contains the factor $\hbar$, its value may be negligible in the classical limit.   Though not an essential part of the postulates of quantum mechanics,  spin is thus considered as a quantum phenomenon. In the classical limit, there would effectively be no spin angular momentum since all the components of a many component wave function reduces to the same function.

\subsection{Expansion postulate: general case}

 As we have seen above, a quantum wave function $\psi({\bf r})=\Psi({\bf r},t=0)$ can be expanded in terms of eigenfunctions of an observable of the system. The eigenvalue equation for such an observable $A_{op}$ in the discrete case  is of the form

\begin{equation}
A_{op} v_{i}({\bf r})=a_i v_{i}({\bf r}). \label{eq:eigen_value_eqn}
\end{equation}
When $A_{op}$ is a Hermitian operator, the eigenvalues $a_i$ shall be real.  (Here we consider only the discrete case. The discussion can be easily extended to the case of continuous eigenvalues.)

The expansion postulate can now be stated  explicitly for the most general case. Let the state of the system  be represented by a normalised wave function $ \Psi({\bf r},t)$.  The expansion of this state in terms of the normalised eigenkets $v_{i}$ of any observable of the system is 

\begin{equation}
\Psi({\bf r},t) =\sum_i c_i (t) v_{i}({\bf r}), \label{eq:expansion_psi_v}
\end{equation}
where $c_i (t)$'s are the appropriate coefficients. As in the earlier cases, $c_i$ can be found using equation (\ref{eq:c_n}) as

\begin{equation}
c_i(t)=\int v_i^{\star}({\bf r}) \Psi({\bf r},t) dV.
\end{equation}

This makes the mathematical formalism of quantum mechanics consistent, in the sense that  any square integrable wave function can be a solution of the Schrodinger equation and it can be expanded in terms of the eigenstates of any  of its observables.

\section{Probability  for any observable} \label{sec:born_any}

We shall now extend Born's probability axiom to obtain a more general form for the same. This axiom, stated earlier in subsection \ref{subsec:born}, refers only to the role of $|\Psi({\bf r},t)|^2dV$ as the probability to find the particle in a volume $dV$ around the point ${\bf r}$ in configuration space.  However, the postulate  can now be extended to a more general statement, applicable to any observable physical quantity. 

First let us consider the expansion of the wave function $\Psi({\bf r},t)$ of the system for the nondegenerate case, as in equation (\ref{eq:unit_evol}). Assuming  this to be a normalised wave function, we  have

\begin{equation}
\int \Psi^{\star}({\bf r},t) \Psi({\bf r},t) dV=\sum_i |c_i|^2 =1.
\end{equation}
 Since the right hand side is a probability, each term in the summation must also  be a probability. 
Thus $|c_i|^2$ may be interpreted as the probability  to obtain the $i^{th}$ energy eigenvalue $E_i$ in a measurement. In the nondegenerate case, this energy value corresponds to the energy eigenstate $\psi_i({\bf r})$ alone. In the degenerate case, we need to sum over the quantity  over all the eigenstates with the same energy value. The corresponding equation can be read as 

\begin{equation}
\sum_i \sum_{k=1}^{g_i} |c^{k_i}|^2 =1.
\end{equation}
It is understood that when the $i^{th}$ energy eigenvalue is $g_i$-fold degenerate, the probability to obtain the energy as $E_i$ in a measurement can be 

$$
\sum_{k=1}^{g_i} |c^{k_i}|^2 .
$$ 

Now we consider  the general case of any observable $A_{op}$, where the expansion postulate is as in equation (\ref{eq:expansion_psi_v}). Here we may postulate that $|c_i(t)|^2$ is the probability to obtain the nondegenerate eigenvalue $a_i$ when the physical quantity $A$ is measured on a system when it is in the normalised state $\Psi$, at time $t$. Also we assume that a measurement of the quantity $A$ is certain to give any one of the eigenvalues $a_i$.

 When the eigenvalues $a_i$ are all  nondegenerate, the postulate of probability can  be extended to all observables by stating that the probability of getting the eigenvalue $a_i$ in the measurement of the observable  $A_{op}$ at time $t$ is  

\begin{equation}
{\cal P}(a_i,t)=|c_i(t)|^2. \label{eq:prob_cn}
\end{equation}   
Making use of the expansion postulate, the total probability to obtain any one of the eigenvalues can be written as

\begin{equation}
\Sigma_i |c_i(t)|^2=1. \label{eq:A}
\end{equation} 
Making use of equation (\ref{eq:c_n}), one can obtain

\begin{equation}
{\cal P}(a_i,t)=|c_i(t)|^2= \left| \int_V v_{i}^{\star}({\bf r})\Psi({\bf r},t)dV \; \right|^2.
\end{equation}

To include the  degenerate case, we consider an eigenvalue $a_i$ that is $g_i$-fold degenerate. This means that there can be $g_i$ linearly independent eigenfunctions corresponding to the eigenvalue $a_i$. The eigenvalue equation for this case is

\begin{equation}
A_{op} v^k_{i}({\bf r})=a_i v^k_{i}({\bf r}), \; \; \; k=1,2,3, ..g_i. \label{eq:eigen_value_eqn_degen}
\end{equation}
The expansion postulate (\ref{eq:expansion_psi_v}) can now be written in a  modified form as

\begin{equation}
\Psi({\bf r},t) =\sum_i \sum_{k=1}^{g_i} c^{k}_i(t) v^k_{i}({\bf r}) . \label{eq:expansion_Psi_degen}
\end{equation}
The postulate of probability now states that the probability for obtaining the eigenvalue $a_i$ in a measurement of $A$ is 

\begin{equation}
{\cal P}(a_i,t)=\sum_{k=1}^{g_i}\; \left| c^{k}_i(t)\right|^2 =\sum_{k=1}^{g_i} \; \left| \int_a^b (v^k_{i})^{\star}({\bf r})\Psi({\bf r},t)dV \; \right|^2. \label{eq:prob_cn_degen}
\end{equation}

Lastly, we state the probability axiom for the case where the spectrum of $A$ is continuous. The eigenvalue equation in this case can be written as

\begin{equation}
A_{op} v({\bf r},a)=a\; v({\bf r},a).
\end{equation}
The expansion postulate now becomes

\begin{equation}
\Psi({\bf r},t)=\int c(a,t)v({\bf r},a) da . \label{eq:expansion_a}
\end{equation}
Since the measured value of the physical quantity is a continuous variable, we must define a probability density $ \rho(a,t)$. Let ${d\cal P}(a)$ be the probability to obtain the measured value of the observable to be between $a$ and $a+da$ at time $t$. Then 

\begin{equation}
d{\cal P}(a,t)=\rho (a,t)da.
\end{equation} 
The postulate of probability in the continuous eigenvalue case is that the probability density (i.e., the probability per unit interval of $
a$) for getting the measured value of $A$ around  $a$, at time $t$, is given by

\begin{equation}
\rho(a,t)=|c(a,t)|^2 = \left| \int_V v^{\star}(a)\Psi({\bf r},t)dV\; \right|^2. \label{eq:prob_density}
\end{equation}
Using the normalisation condition, it is possible to show that the integral

\begin{equation}
\int \rho(a,t)da = \int |c(a,t)|^2da =1.
\end{equation}

\subsubsection{Eigenstate of position}
   
   We have postulated that the position operator  ${\bf r}_{op}$ is  Hermitian and that its eigenstates $w$  have the completeness property. Thus we expect to expand any wave function $\psi({\bf r})$ in terms of the position eigenfunctions $w({\bf r},{\bf r}^{\prime})$; i.e.,
   
   \begin{equation}
   \psi({\bf r})=\int_V c({\bf r}^{\prime})w ({\bf r},{\bf r}^{\prime}) dV^{\prime}. \label{eq:expansion_x}
\end{equation}    
 We shall now find an explicit form for $w({\bf r},{\bf r}^{\prime})$  by considering the role of  $c({\bf r}^{\prime})$ in this equation (\ref{eq:expansion_x}).  In this case, $|c({\bf r}^{\prime})|^2$ must be the probability density  to find the particle around the point ${\bf r}^{\prime}$. When combined with Born's probability axiom, we conclude that $c({\bf r}^{\prime})$ must be identical to $\psi({\bf r}^{\prime})$. With the definition of Dirac $\delta$-function given by equation (\ref{eq:deltafn}), we conclude from the above equation that $w ({\bf r},{\bf r}^{\prime})$ must be the Dirac delta function. i.e.,

\begin{equation}
w({\bf r},{\bf r}^{\prime})=\delta({\bf r}-{\bf r}^{\prime}). \label{eq:pos_eig_fn}
\end{equation}

Thus  the probability axiom helps us to identify  the position eigenfunction as the  Dirac delta function.

\section{Observable fields} \label{sec:mom_energy_fields}
In this section, we deviate from the standard formulation of quantum mechanics and introduce the concept of observable fields, such as momentum field and energy field.  In classical mechanics, there exist a method  to find the solution of the mechanical problem using the HJ formalism, as discussed in subsection \ref{subsec:soln_mech_HJ_alt}. Here  we need only to integrate the classical equation of motion (\ref{eq:canon2_1_2}) for obtaining the trajectories. In three dimensions, the equation of motion is given by (\ref{eq:eqn_of_motion}). In  subsection \ref{subsec:mom_oper}, we have taken a special  (free particle) case of the classical equation of motion (\ref{eq:eqn_of_motion}) to obtain (\ref{eq:mom_eqn}), the eigenvalue equation for momentum. (This helped us to identify the momentum operator, whose eigenfunctions are the plane waves.) We can use an equation of the same form (\ref{eq:eqn_of_motion}) also in quantum mechanics and use it as an equation of motion to obtain quantum trajectories. For a one particle case, this quantum equation of motion is

\begin{equation}
{\bf p}({\bf r},t)=-i\hbar \frac{1}{\Psi({\bf r},t)}\nabla \Psi({\bf r},t) =\frac{{\bf p}_{op} \Psi({\bf r},t)}{\Psi({\bf r},t)}. \label{eq:mom_field}
\end{equation} 
Here, one must take care of the fact that this  gives the canonical momentum and not the mechanical momentum. Once we obtain ${\bf p}$ in terms of $\dot{\bf r}$, we can integrate this with respect to $t$ to obtain trajectories. It may be noted that the trajectories we obtain in this manner are  in general  complex trajectories \cite{mvj}. 

Another salient feature of the above expression is that it gives the momentum field ${\bf p}$, and it is defined over the configuration space of the system. A similar definition for the energy field can also be given as 

\begin{equation}
E({\bf r},t)=\frac{H_{op}\Psi({\bf r},t)}{\Psi({\bf r},t)}.\label{eq:en_field1}
\end{equation} 
In the case of a one-particle case, we write it more explicitly as

\begin{equation}
E({\bf r},t)= -\frac{\hbar^2}{2m}\frac{1}{\Psi({\bf r},t)} \nabla^2\Psi({\bf r},t)+V({\bf r},t) \label{eq:en_field2}
\end{equation}

Similar fields can be defined for other physical quantities as well. A general expression for the field $A({\bf r})$ corresponding to any observable $A_{op}$ is
 
\begin{equation}
A({\bf r},t)=\frac{A_{op}\Psi({\bf r},t)}{\Psi({\bf r},t)}. \label{eq:A_field}
\end{equation} 
Note that in all cases, $\Psi({\bf r},t)$ is the  wave function for the system at the given time, defined over the configuration space. It is easy to see that if the state of the system is an eigenstate of an  operator, then the corresponding  field will be a constant, equal to its eigenvalue, throughout all configuration space. But this is only for that particular observable; the fields of  all other physical quantities in this case may be variable.   Specifically, when $\Psi$ is a superposition of energy eigenstates, the energy field will be variable, depending on both ${\bf r}$ and $t$. But for an eigenstate of energy, the energy field will be a constant, as can be seen from equation (\ref{eq:en_field2}).

 We shall find below that such fields are useful for computing expectation values of corresponding physical quantities, resorting only to the standard definition of expectation values  in probability theory. 

 \subsection{Classical limit of fields}
 It was seen that in the limit $\hbar \rightarrow 0$, the quantum HJ equation (\ref{eq:mod_HJ_eqn}) reduces to the classical HJ equation. Consequently, in the classical limit, the  Schrodinger equation (\ref{eq:SE})  reduces to the classical mechanics wave equation  (\ref{eq:class_waveqn}). It is only natural to expect that the classical limit of the quantum wave function $\Psi({\bf r},t)$ is the classical mechanics wave function $X({\bf r},t)$. In the  classical case also, we continue to define the fields as above. Then for an operator $A_{op}$, the same expression (\ref{eq:A_field}) holds, with $\Psi({\bf r},t)$ replaced by $X({\bf r},t)$.  The classical  expression for the energy field is of the same form  as that in equation (\ref{eq:en_field1}), with 
 
 \begin{equation}
 E({\bf r},t)=\frac{ H_{op}X({\bf r},t)}{X({\bf r},t)} \equiv -\frac{\hbar^2}{2m}\frac{1}{X^2({\bf r},t)}[\nabla X({\bf r},t)]^2+V({\bf r},t).
 \end{equation}
Here $H_{op}$ is the classical Hamiltonian operator, which can be deduced from equation (\ref{eq:class_waveqn}). Note that this is not  a linear operator. When the Hamiltonian is independent of time $t$, one can define the energy field for the classical case as
 
 \begin{equation}
 E=-\frac{\hbar^2}{2m}\frac{1}{X^2({\bf r},t)}[ \nabla X({\bf r},t)]^2+V({\bf r}),
 \end{equation}
In this case,  since the solutions $X({\bf r},t)$ are eigenstates of the classical $H_{op}$,  the energy field will  be a constant everywhere. But other  observables will have their fields varying with position.

\section{Measurement of physical quantities}
The measurement postulate in quantum mechanics states that the only measurable values of a physical observable are the various eigenvalues of the corresponding observable. When the measurement of a physical quantity  made on the system in a state $\Psi({\bf r},t)$ gives the result $a_i$, the state of the system immediately after the measurement shall be $v_i({\bf r})$, an eigenfunction corresponding to the value $a_i$ of the observable.

\subsection{Expectation values}
 
 The expectation value of any physical quantity can be evaluated using the wave function $\Psi$  of the system. Let $A_{op}$ is the operator corresponding to a physical quantity $A$. In quantum mechanics, it is  postulated  that the expectation value of $A$ in a measurement is

  \begin{equation}
\langle {A} \rangle = \int_V \Psi^{\star}({\bf r},t)\; {A}_{op}\; \Psi({\bf r},t) dV. \label{eq:expctn_A}
\end{equation}

We have seen earlier in subsection \ref{subsec:born} that the probability axiom can be used to evaluate the expectation value of the position of a particle during some position measurement.  It follows from the postulate of probability  in quantum mechanics that the expectation value of  the momentum of the particle ${\bf p}$ is 
 
 \begin{equation}
\langle {\bf p} \rangle = \int_V \Psi^{\star}({\bf r},t)\; {\bf p}_{op}\; \Psi({\bf r},t) dV = - i\hbar \int_V \Psi^{\star}({\bf r},t)\; \nabla \Psi({\bf r},t) dV, \label{eq:expctn_mom}
\end{equation}
where we have used ${\bf p}_{op}=-i\hbar \nabla$.
 Similarly,  the expectation value of the energy of the particle $E$ can be evaluated  as
  \begin{equation}
\langle {E} \rangle = \int_V \Psi^{\star}({\bf r},t)\; {H}_{op}\; \Psi({\bf r},t) dV. \label{eq:expctn_energy}
\end{equation}

If the the operator corresponding to the  position variable of a particle is ${\bf r}\equiv {\bf r}_{op}$ itself, then equation (\ref{eq:expctn_position1})  can  equivalently be written as

\begin{equation}
\langle {\bf r} \rangle = \int_V \Psi^{\star}({\bf r},t)\; {\bf r}_{op}\; \Psi({\bf r},t) dV, \label{eq:expctn_position2}
\end{equation}
which is in conformity with the expressions for expectation values of momentum and energy. It should be noted that the expression (\ref{eq:expctn_A}), which gives the expectation values of observables, is only a postulate in quantum mechanics.

\subsection{A more fundamental expression for expectation values}
In section \ref{sec:mom_energy_fields}, we have defined the momentum and energy fields for a system as function of position and time. This concept was later extended to any observable. We have defined the field corresponding to an observable $A$ as

\begin{equation}
A({\bf r},t)=\frac{1}{\Psi({\bf r},t)}A_{op}\Psi({\bf r},t).
\end{equation}
With the help of this definition, one can write the expectation values of any observable as 

\begin{equation}
\langle {A} \rangle = \int_V A({\bf r},t) \Psi^{\star}({\bf r},t) \Psi({\bf r},t) dV.
\end{equation}
For example, it is easy to see that the mean value of momentum can be written as an average of the momentum field ${\bf p}({\bf r},t)$ taken over all space, with $\Psi^{\star}({\bf r},t) \Psi({\bf r},t)$ as probability densities.

\begin{equation}
\langle {{\bf p}} \rangle = \int_V {\bf p}({\bf r},t) \Psi^{\star}({\bf r},t) \Psi({\bf r},t)dV
\end{equation}
This can be seen to yield the same expression in equation (\ref{eq:expctn_mom}). Similarly, the mean value of energy can be computed in this alternative approach as
\begin{equation}
\langle {E} \rangle = \int_V E({\bf r},t) \Psi^{\star}({\bf r},t) \Psi({\bf r},t) dV,
\end{equation}
which is the same as that in equation (\ref{eq:expctn_energy}).

We note that the concept of fields associated with observables is useful in this context. The expression for the expectation values are in conformity with the standard definition of probability, with the corresponding fields as random variables.

\section{Summary and discussion}

de Broglie's principle of wave-particle duality is the first among the revolutionary ideas that led  to the development of modern quantum mechanics. This duality principle was the crystallisation of thoughts on the fundamental nature of light, which behaved  like both waves and  particles. The wave nature of light was well-established through the electromagnetic theory, and its particle-like nature  became evident through the explanation of  photoelectric effect and the Compton effect. Once the duality principle is applied, it became imperative that  matter particles like electrons, which  are to be described by waves, should obey an appropriate  wave equation. Such a wave equation for particles was not available until Schrodinger came up with his equation in 1926. Schrodinger made these attempts based upon the framework of classical  theory of waves, rather than classical mechanics per se. 

As is well-known, the matrix mechanics formulation of Heisenberg, made  in the year 1925,  preceded the work of Schrodinger. Heisenberg did not at all use the idea of a wave in his formalism, yet he could obtain almost the same results that Schrodinger obtained through his wave mechanics.   However, it may be noted that one can  easily  get  matrix mechanics from wave mechanics. Evaluation of matrix elements, as in matrix mechanics, is there in the final step while solving problems in wave mechanics. But the opposite route of obtaining waves from matrices was not appealing to Heisenberg because he thought that  the quantum wave is merely a mathematical tool for calculations and that it has no real existence. Moreover, he considered the quantum state of the system as merely subjective,  representing only our state of knowledge about the system. Recently, the issue of objective reality of wave function (i.e., its existence independent of observers) has arisen as a hot topic. This is in view of the new experiments which claim to establish that quantum states are real and not merely our state of knowledge or information about the system.

 In classical mechanics, the position and momentum values of all particles in a many-particle system specifies the state of the system whereas in quantum mechanics, the state of the system is postulated to be represented by the wave function $\Psi$. In this treatise, we introduced a new element to the conventional classical mechanical viewpoint  with regard to the definition of the `state of a system'.   We have chosen  $X({\bf r},t)=\exp(iS({\bf r},t)/\hbar)$  (with $S({\bf r},t)$ as the solution of the classical HJ equation)  as representing the classical `energy state' of the system and named it as the the classical mechanics wave function. This approach helped us to  develop Schrodinger's wave mechanics as a natural outgrowth of classical mechanics. 
 
 Our first step was to review how the classical eikonal equation for geometrical light propagation is obtained from Maxwell's electromagnetic wave equations, as a limiting case. The major distinction between these two is that the latter admits superposition of its own solutions whereas the former does not have such a freedom.   We now made a comparison of this with the classical HJ equation or the equivalent classical mechanics wave equation and noted that they are  nonlinear and do not allow superposition of solutions. The implication is that whereas the photon wave function can be any square integrable function and remain a solution of Maxwell equations, the classical mechanics wave function can take only certain specific form while being the solution of the (nonlinear) classical mechanics wave equation. We have suggested that de Broglie's wave-particle duality must be extended in such a way that both wave and particle  states can be represented by any function  belonging to the set of all square integrable functions, while being solutions of their respective fundamental equations.  It was noted that one needs only to add a term of the order of $\hbar$ to the classical HJ equation  to make the  equivalent  classical mechanics wave equation linear and as one allowing the superposition principle. The result is the Schrodinger equation in quantum mechanics. This is demonstrated as an alternative approach for deducing the Schrodinger equation.

If one had arrived at the Schrodinger formalism along this route, starting from the classical mechanics wave equation, the interpretation of quantum mechanics would not have appeared so bizarre. When written for a free particle, the equation of motion in  the classical HJ theory  assumes the form of an eigenvalue equation for momentum. Obviously, functions such as the classical mechanics wave function can  be expanded into a series in terms of the complete set of eigenfunctions of this momentum operator. This, in fact, is the Fourier series expansion for the energy state of the system. Also we note that  the quantum probability rule can  be applied also to the classical mechanics wave function $X ({\bf r},t)\equiv e^{iS({\bf r},t)/\hbar}$. Using this in any mechanical problem, we can obtain a normalised probability distribution. Since the action $S$ is real in the classical case, the  probability density can be found a constant in the classically allowed region. Beyond that region, the probability distribution function falls very sharply to zero. When this wave function is expanded in terms of momentum eigenfunction (the Fourier series), one can assign a  probability  for each momentum eigenvalue. For instance, in the classical discrete case, it is possible to  assign the probability for the $n^{th}$ eigenvalue  as $\mid c_n\mid ^2$, the modulus square of the  coefficient of the term corresponding to this momentum.  A similar statement can be made for any other observable with complete set of eigenfunctions too.  But now the question arises whether this  quantity is the probability to obtain a value of the observable when a measurement of it is made on the system. As we see below, such an interpretation  cannot be made in the classical case, since the phenomenon of wave function collapse  cannot occur there.

The collapse of the wave function of a system means a transition of its energy state  into an eigenstate of an observable   under a measurement. One can see that this is a purely quantum mechanical effect, and is possible because eigenstates of an observable is also an energy state, due to the superposition principle. On the other hand,  he classical energy state of the system, described by the  classical mechanics wave function, can take only specific forms (eigenstates of classical Hamiltonian operator)  and an observation will not make the wave function  collapse to an eigenstate of the measured observable. This is because such eigenstates cannot in general be solutions of the classical mechanics wave equation.   In all classical cases, a wave function collapse to an eigenfunction of another observable cannot happen. Hence in such cases, probabilities such as $\mid c_n \mid^2$ have no physically meaningful interpretation.  Classically, a measurement can at most make a transition from one energy state to another energy state. This would involve some energy transfer between the system and the measuring device, which can be termed a physical interaction between the two.    We realise that for quantum systems too, a collapse of the wave function can occur only when there is a physical change that affects the state of the system. The same thing can happen during  measurements. Viewed in this way, one can  conclude that the wave function collapse and the wave function itself are real phenomena and not merely   subjective `state of knowledge' of the observer.

In this context, it is easy to see why the phenomenon of entanglement does not occur in the classical realm. One cannot speak meaningfully about entanglement when there is no wave function collapse, as it happens in classical mechanics. 

Another feature that is said to have no classical analogue is the intrinsic spin of particles. But we have seen that the intrinsic spin can be understood as arising due to  many-component wave functions of such particles. For particles with only one component for their wave function, there shall be no  intrinsic spin. Even for cases with more than one component, we have seen that in the limit when $\hbar$ is negligible, there is effectively  only one component and hence the intrinsic spin will not become explicit.  Hence one can understand that in the classical limit, the absence of intrinsic spin is nothing other than the absence of more-than-one distinct component wave function for the particle.

Another notable feature  we find in both classical and quantum realm is the presence of fields related to each observable of a system, such as the energy field, momentum field, etc. These fields are  directly related to the  wave functions. Our expression for expectation value of these observables in both realm, with $\mid \Psi \mid^2$ taken as the position probability density and these fields  taken as the random variables,  is the standard one in probability theory. Thus the above-mentioned expression for expectation value can be considered ontological, i.e., it depends on  physically existing fields.   This, again, endorses the reality of these fields and in turn, the reality of the quantum wave function itself. 

 We thus have seen that  what really distinguishes quantum mechanics from classical mechanics is the validity of the principle of superposition. As a result of obeying this principle, in quantum mechanics,   energy states can be described by any function belonging to  the set of square-integrable functions and the eigenfunction of any observable can be expanded in terms of eigenfunctions of other observables, including that of the Hamiltonian. But the representation of the energy state of a system with wave functions, the series expansion of such state functions in terms of complete set of eigenfunctions of  observables other than that of energy, the assignment of probability for each eigenvalue of these observables, etc. can clearly be defined for the classical case too. In classical mechanics, eigenfunctions of other observables are not possible states of the system, since an expansion of them in terms of energy states is not possible.  It is only for this reason that  collapse of the wave function,  entanglement between parts of the system,  etc., are not observed in the classical limit.      Most of the puzzles in quantum mechanics originate from these  features, but we see that they have their seeds in classical mechanics too. It is rightfully expected that those puzzles  fade away while one takes this proper route from classical mechanics to quantum mechanics.

 \end{document}